\documentclass[a4paper, 10pt, conference]{IEEEtran}      % Use this line for a4 paper

\IEEEoverridecommandlockouts                              % This command is only needed if 
\usepackage{amsmath}
\usepackage{graphicx}
\usepackage{svg}
\usepackage[colorlinks=true, allcolors=blue]{hyperref}

\usepackage{cite}
\usepackage{multirow}
\usepackage{adjustbox}
\usepackage{amsmath,amssymb,amsfonts}
\usepackage{graphicx}
\usepackage{textcomp}
\usepackage{xcolor}
\usepackage{xcolor,soul}
\usepackage{algorithm}
\usepackage{algpseudocode}

\usepackage{times} % assumes a new font selection scheme installed
\usepackage{amsmath} % assumes amsmath package installed
\usepackage{amssymb}  % assumes amsmath package installed

\usepackage{graphicx}  % Written by David Carlisle and Sebastian Rahtz
\usepackage{amsmath}   % From the American Mathematical Society
\usepackage{array}
\usepackage{times}
\usepackage{comment}
\usepackage{latexsym,theorem}
\usepackage{amsmath,amsfonts,amssymb,amsbsy}
\usepackage{subfig}
\usepackage{tikz}

\definecolor{yellowcolor}{HTML}{ffff00}
\definecolor{bluecolor}{HTML}{a6cee8}

\def\BibTeX{{\rm B\kern-.05em{\sc i\kern-.025em b}\kern-.08em
    T\kern-.1667em\lower.7ex\hbox{E}\kern-.125emX}}

{\theorembodyfont{\slshape}\newtheorem{theorem}{Theorem}[section]}
{\theorembodyfont{\slshape}}
{\theorembodyfont{\slshape}}
{\theorembodyfont{\slshape}}
{\theorembodyfont{\upshape}\newtheorem{definition}[theorem]{Definition}}
{\theorembodyfont{\upshape}}
{\theorembodyfont{\upshape}\newtheorem{remark}[theorem]{Remark}}
{\theorembodyfont{\upshape}\newtheorem{assumption}[theorem]{Assumption}}

\begin{document}

\title{Multivariable control of modular multilevel converters with convergence and safety guarantees }

\author{Victor Daniel Reyes Dreke$^{1}$, Ygor Pereira Marca$^{2}$,  Maurice Roes$^{3}$ and Mircea Lazar$^{4}$% <-this % stops a space
\thanks{This research was funded by the NEON (New Energy and Mobility Outlook for the Netherlands) Crossover NWO (Dutch Research Council) Grant, project number 17628.}% <-this % stops a space
\thanks{Affiliation: $^{[1,4]}$ Control Systems Group, Department of Electrical Engineering, Eindhoven University of Technology, The Netherlands. $^{[2,3]}$ Electromechanics and Power Electronics Group, Department of Electrical Engineering, Eindhoven University of Technology, The Netherlands. E-mails: {\tt\small $^{1}$ v.d.reyes.dreke@tue.nl, $^{4}$ m.lazar@tue.nl.}} %
}
\maketitle

\begin{abstract}
Well-designed current control is a key factor in ensuring the efficient and safe operation of modular multilevel converters (MMCs). Even though this control problem involves multiple control objectives, conventional current control schemes are comprised of independently designed decoupled controllers, e.g., proportional-integral (PI) or proportional-resonant (PR). Due to the bilinearity of the MMC dynamics, tuning PI and PR controllers so that good performance and constraint satisfaction are guaranteed is quite challenging. This challenge becomes more relevant in an AC/AC MMC configuration due to the complexity of tracking the single-phase sinusoidal components of the MMC output. In this paper, we propose a method to design a multivariable controller, i.e., a static feedback gain, to regulate the MMC currents. We use a physics-informed transformation to model the MMC dynamics linearly and synthesise the proposed controller. We use this linear model to formulate a linear matrix inequality that computes a feedback gain that guarantees safe and effective operation, including (i) limited tracking error, (ii) stability, and (iii) meeting all constraints. To test the efficacy of our method, we examine its performance in a direct AC/AC MMC simulated in Simulink/PLECS and in a scaled-down AC/AC MMC prototype to investigate the ultra-fast charging of electric vehicles.
\end{abstract}
% \begin{key}
% Distributed Model Predictive Control, Stability, Recursive feasibility, Synchronization, Multi-Agent Systems
% \end{IEEEkeywords}
% ===================================================================   Introduction    ===================================================================
\section{Introduction}
\label{sec:introduction}

Modular multilevel converters (MMCs) are novel voltage source converters that offer high efficiency, scalability, and low harmonic distortion \cite{Marquardt2003MMC,Perez2021MMCSurvey, Challa2023MMCSurvey, Marquardt2018MMCSurvey}.
Well-designed current control is a key factor in ensuring their safe and efficient operation \cite{Du2018MMCBook}.
This control problem involves multiple control objectives, i.e., regulating the grid current, the MMC output current, and the sum of the capacitor voltages in each arm.
\textcolor{black}{In this case, regulating both currents controls the power transmitted from the grid to the MMC output.}
Furthermore, the safe operation depends mainly on stabilising the sum of the capacitor voltages of each arm, i.e., \textcolor{black}{total arm voltage}\cite{Sharifabadi2016MMCBook}.

The conventional control schemes (CCSs) of the MMC currents, as defined in \cite{Du2018MMCBook} and \cite{Sharifabadi2016MMCBook}, consist of a set of independently designed controllers for each control objective.
In this case, these controllers are usually single-input, single-output (SISO) controllers, such as proportional-integral (PI) or proportional-resonant (PR) controllers, which are not designed to deal with constrained multivariable control problems.
As MMCs are multivariable systems with bilinear dynamics and constrained control inputs, a major issue in the design of CCSs is tuning the controllers to ensure the desired performance, stability, and constraint satisfaction.
Commonly, these tunning procedures are based on trial-and-error methods.
As a consequence, when compared with multivariable control solutions as linear quadratic regulator (LQR) or model predictive control (MPC), the CCS's performance is lower ( see e.g. \cite{Dekka2019MMCSurvey,Böcker2015MMC,Fuchs2020LQR,Darivianakis2014MMCMPCSurvey,ReyesDreke2022NMPC}).

\textcolor{black}{Furthermore, the complexity of the CCS depends on the waveform shape of the MMC output.}
\textcolor{black}{For MMCs operating under AC outputs, i.e., AC/AC MMC, this complexity increases due to the challenges of tracking single-phase AC signals.}
\textcolor{black}{A usual solution, i.e., PI, can stabilize sinusoidal signals as in \cite{Pereira2021ExampleACAC,Song2018MMCACAC,YU2023MMCACAC}, but they can not achieve zero error tracking.} 
\textcolor{black}{The alternative, i.e., PR, has a narrow frequency band around the resonance frequency.}
\textcolor{black}{Hence, perturbations in the resonance frequency can reduce the controller performance dramatically \cite{Timbus2006MMCPR}.}

Guaranteeing convergence of the controlled variables while meeting constraints is another of the challenges of the CCS.
In \cite{Harnefors2015MMC17}, it is shown that using PI controllers can ensure MMC's safe operation, i.e., having stable total arm voltage.
In this case, unconstrained continuous-time PIs used in AC/DC MMCs are considered.
\textcolor{black}{However}, \cite{Harnefors2015MMC17} does not discuss tuning methods or performance of the controllers.
An alternative approach is multivariable control, i.e., state feedback gain \cite{Zama2021MMCRobust,Tavakoli2022MMCRobust}.
These solutions also use a linear model of the MMC in the dq-reference framework.
Despite having better performance and being robust against model uncertainty, they are unconstrained and only compatible with AC/DC MMCs.
Nowadays, MPC is the only implemented solution that offers high performance, convergence, and constraint satisfaction \cite{Challa2023MMCSurvey}.
However, its use is minimal due to its computational complexity, especially in AC/AC MMCs \cite{Du2018MMCBook}.

In this paper, we propose a framework to design a multivariable controller, i.e., static feedback gain, to control the MMC currents in the $abc$-reference framework regardless of the configuration.
We use a physics-informed transformation to model the MMC dynamics linearly and synthesise the proposed controller.
Using linear matrix inequalities (LMIs) inspired by \cite[Theorem 1 c.f.]{Francis1976LOT}, we synthesise a controller that, in closed-loop with the linear model, reaches zero-error tracking while recursively meeting constraints inside a level set.
This level set, i.e., an invariant domain of attraction, is a byproduct of the LMI.
% Moreover, we show the suitability of the controller by proving that the tracking error of the MMC currents is bounded while keeping the total arm voltage stable, i.e., safe operation.
% Moreover, we prove that, in a closed loop with the MMC, the proposed controller guarantees well-designed tracking of the MMC currents, i.e., safe and stable operation and bounded tracking error
Moreover, we prove that, in a closed loop with the MMC, this controller guarantees a bounded tracking error of the MMC currents while keeping the total arm voltage stable, i.e., safe operation.
We demonstrate the controller efficacy in two scenarios: (i) simulation in a Simulink/PLECs environment and (ii) a scaled-down MMC prototype of direct AC/AC MMC for ultra-fast chargers.

% The outline of the paper is as follows: Section\ref{sec:mmcModelling} and~\ref{sec:mmcControlProblem} introduce the average model and the control problem of the MMC. 
% The methodology for the synthesis of the feedback gain controller is presented in Section~\ref{sec:mmcCurrentController}.
% The performance of the proposed controller is tested in a simulated and scaled-down prototype; these results are presented in Section~\ref{sec:mmcSimulationResults} and Section~\ref{sec:ExperimentalVerification}.
% Finally, conclusions are shown in Section~\ref{sec:Conclusion}.

\subsection{Basic Notation:}
Let $\mathbb{N}$, $\mathbb{N}_{+}$, $\mathbb{R}$, $\mathbb{R}_{{+}}$, and $\mathbb{Z}$ denote the set of natural numbers, the set of natural numbers \textcolor{black}{excluding} 0, the field of real numbers, the field of non{-}negative real numbers and the set of integer numbers, respectively. 
For a set $\mathbb{S}\subseteq\mathbb{R}^n$ define $\mathbb{S}_{[a,b]}:=\{s\in\mathbb{S} \ : \ a\leq s\leq b\}$. 
Define an $n${-}dimensional vector filled with ones as $\mathbf{1}_{n}$ and one filled with zeros as $\mathbf{0}_{n}$. 
The identity matrix is denoted as $\mathbf{I}_{n}$. 
A block diagonal matrix with matrices $A_1$ to $A_m$ on the diagonal is denoted by $\operatorname{diag}\left(A_1, {...}, A_m\right)$. 
For matrices $A, B \in \mathbb{R}^{n\times m}$, horizontal and vertical concatenation are defined as $[A,B] {=}  [A \; B]$ and $[A;B]{=}[A^\top,B^\top]^\top$, respectively.

% The operator $\otimes$ is the standard Kronecker product. 
We write $A \succ 0 \;(A \succcurlyeq 0)$ for a symmetric, positive (semi)definite matrix $A=A^{\top} {\in} \mathbb{R}^{n \times n}$. 
The $i$ eigenvalues of a matrix $A{\in} \mathbb{R}^{n \times n}$ are denoted by $\lambda_i(A)$ and the spectral radius of $A$ is $\rho(A) := \max{\{|\lambda_1{(A)}|, {...}, |\lambda_n(A)|\}}$.
A discrete-time state-space model, i.e., $x(k+1) = Ax(k) + B u(k)$, is represented as $x^+ =A x + B u$, where $x^+ := x(k+1)$ and $x := x(k)$.
% ============================================================================================================================================
% =======================================================  MMC Average Modelling  ============================================================
% ============================================================================================================================================

\section{MMC Average Modelling}
\label{sec:mmcModelling}

\begin{figure}[!t]
	\centering
	\includegraphics[width=0.9\columnwidth]{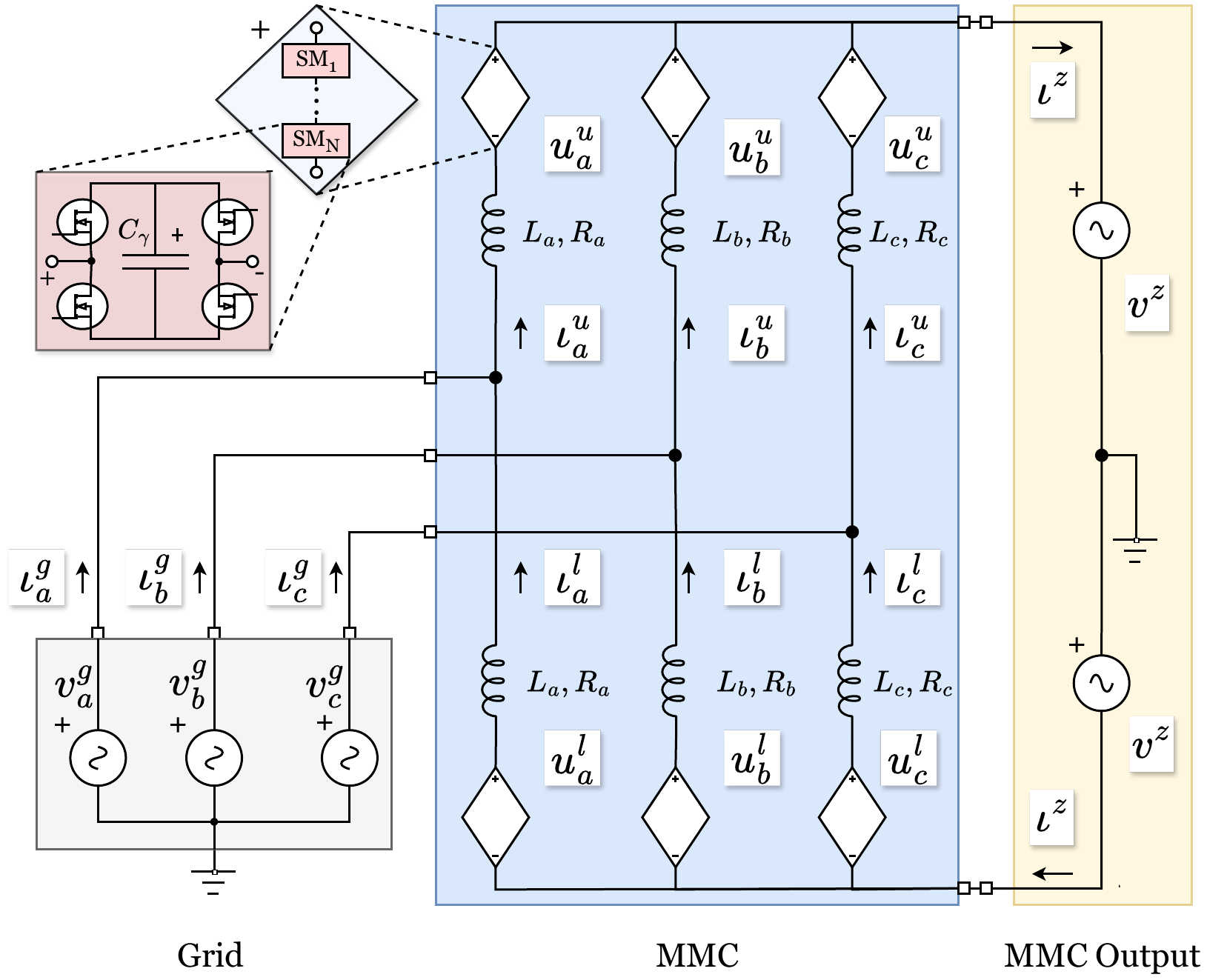}
	\caption{\textcolor{black}{Schematic of a three-phase bidirectional direct AC/AC MMC topology with full-bridge submodules as presented in \cite{Pereira2021ExampleACAC}.}}
	\label{Fig:MMCDiagram}
\end{figure}

A standard MMC is comprised of three-phase legs, labelled by $m {\in} \{a, b, c\}$, \textcolor{black}{each of which consists} of two arms, labelled by $n {\in} \{u,l\}$ corresponding to the upper and lower arm, respectively.
The arms are composed of $N$ modules (SM) connected in series, an arm inductor ($L_m$), and an arm resistance ($R_m$) that models the conduction losses of the arm.
This paper considers an MMC equivalent circuit as in \cite{Pereira2021ExampleACDC, Pereira2021ExampleACAC}, see  Figure~\ref{Fig:MMCDiagram}.  
The arm current dynamics, i.e., $\iota^n_m$, for all phases $m{\in} \{a, b, c\}$ yields
\begin{subequations}
 \label{eq:DifferentialEquationUpper}
    \begin{align}
            \label{eq:DifferentialEquationUpper1}
            & L_m \frac{\mathrm{d}}{\mathrm{d} t} \iota_m^{{u}}+R_m \iota_m^{{u}}=u_m^{{u}}+v^g_m-v^{{z}}, \\
             \label{eq:DifferentialEquationUpper2}
            & L_m \frac{\mathrm{d}}{\mathrm{d} t} \iota_m^{l}+R_m \iota_m^{l}=u_m^{l}-v^g_m-v^{{z}},            
    \end{align}
\end{subequations}
with 
\begin{subequations}
\label{eq:nonlinearArmVoltage}
\begin{align}
    \label{eq:nonlinearArmVoltage1}
    u^n_m &=  \sum^{N}_{i=1} s^{\{n,m\}}_i \cdot v^{\{n,m\}}_{C_i}, \;\; \forall n \in \{u,l\}, \\
    \label{eq:nonlinearArmVoltage2}
   \frac{\mathrm{d}}{\mathrm{d} t} v^{\{n,m\}}_{C_i} &= -\frac{1}{C_{\gamma}}s^{\{n,m\}}_i \cdot {\iota}^n_m, \;\; \forall i \in \{1,{...},N\},
\end{align}    
\end{subequations}
where $u^n_m {\in} \mathbb{R}$ is the arm voltage, $C_{\gamma} {\in} \mathbb{R}_+$ is capacitance of the module, and for each $i$ module,  $s^{\{n,m\}}_i {\in} \mathbb{Z}_{[-1,1]}$ is the switching signal, $v^{\{n,m\}}_{C_i} {\in} \mathbb{R_{+}}$ is the capacitor voltage, and $v^z$ and $v^g_m$ are the output and the grid voltage, respectively.
% Note that the transformer voltage waveform depends on the MMC configuration, e.g., AC/AC or AC/DC.
\textcolor{black}{In this case, we can represent the grid and the MMC output currents as a linear combination of the arm currents, i.e.,}
\begin{equation}
   \label{eq:outputCurrentsDefinition1}
       \iota^g_m = \iota^u_m - \iota^l_m \quad \text{and} \quad \iota^z =  \sum_{m=\{a,b,c\}} \iota^z_m, 
\end{equation}
\textcolor{black}{where, for all $m \in \{a,b,c\}$, $\iota^g_m$ are the grid currents and $\iota^z_m := {(\iota^u_m + \iota^l_m)}/{2}$ are the MMC output currents per phase.}
% \textcolor{black}{As in \cite{Pereira2021ExampleACDC}, the grid and the output currents from \eqref{eq:outputCurrentsDefinition1} correspond to the differential and common mode, respectively.}

The complexity and dimension of the model resulting from using \eqref{eq:DifferentialEquationUpper} and \eqref{eq:nonlinearArmVoltage} are high due to the bilinear influence of the integer control signal  $(s^{\{n,m\}}_i)$. 
An accepted solution to reduce the model complexity and dimension is using the model approach as in \cite{Lopez2014MMC, Rohner2011MMCModelling}. 
In these cases, it is assumed that the capacitor voltages are balanced in each arm $n{\in}\{u,l\}$  for all the phases $m {\in} \{a,b,c\}$, such that
\begin{equation}
\label{eq:balancedCapcitorDefinition}
    % \begin{align}
      v^{\{n,m\}}_{C_i}  = v^{\{n,m\}}_{C_j}, \;\; \forall (i,j)\in \{1, ..., N\}.\\  
      % v^{\{l,m\}}_{C_i}  &= v^{\{l,m\}}_{C_j}.
    % \end{align}
\end{equation}    
Consequently, the arm voltage, i.e., $u^n_m$, yields
\begin{equation}
\label{eq:armVoltageLinearDef}
    u^n_m =   \eta^n_m  \mathrm{v}^{n}_{m}, \;\; \forall n \in \{u,l\}
\end{equation}
with 
\begin{equation}
    \mathrm{v}^{n}_{m} {= }\sum^{N}_{i=1} v^{\{n,m\}}_{C_i} , \;\, \text{and} \;\;  \eta^n_m {=}  \frac{1}{N} \sum^{N}_{i=1} s^{\{n,m\}}_i,
\end{equation}
where $\eta^n_m {\in} \mathbb{R}_{[-1,1]}$ is the insertion index and $\mathrm{v}^{n}_{m}{\in} \mathbb{R}$ is \textcolor{black}{the total arm voltage.} 
% \textcolor{black}{the sum of the capacitor voltages of each arm, i.e, total arm voltage.} 
As in \cite{Rohner2011MMCModelling}, we approximate the insertion index to a continuous signal to remove the discontinuities.

Based on \eqref{eq:DifferentialEquationUpper} and \eqref{eq:armVoltageLinearDef}, let us define   a state vector, i.e., ${x} = [x_a, \, x_b, \, x_c]^\top{\in }\mathbb{R}^{12}$, output vector, i.e., $y=[y_a, \, y_b, \, y_c]^\top {\in }\mathbb{R}^{6}$, insertion index vector, i.e., $\eta =[\eta_a,  \, \eta_b,  \, \eta_c]^\top{\in }\mathbb{R}^{6}$, and exogenous input vector, i.e., $w=[w_a,  \, w_b,  \, w_c, \, w_z]^\top{\in} \mathbb{R}^{8}$; where for each phase $m \in \{a,b,c\}$, they are defined as
\begin{subequations}
    \begin{align}
        x_m &{=}\left[{\iota^u_m} \;\, {\iota^l_m} \;\, \mathrm{v}^u_m \;\, \mathrm{v}^l_m\right], 
        y_m {=}\left[\begin{array}{cc} {\iota}^g_m &  {\iota}^z_m \end{array}\right],\\ 
        \eta_m &{=}\left[{\eta^u_m} \;\, {\eta^l_m}\right], 
        w_m {=}\left[\begin{array}{cccc} v^g_m & v^{g'}_m\end{array}\right],
    \end{align}
\end{subequations} 
and 
\begin{equation}
     w_z {=}\left[\begin{array}{cccc}  v^z & v^{z'} \end{array}\right].
\end{equation}
In this case,  ${\iota}^g_m$ and ${\iota}^z_m$ are the grid current and the output current, and $v^{g'}_m$ and $v^{z'}_m$ are obtained from the $90^{\circ}$ phase delay of $v^g_m$ and $v^z_m$, respectively.
The resulting MMC bilinear average model yields a discrete-time state-space model as follows
\begin{equation}
\label{eq:mmcBilinearModelGlobal}
    \begin{aligned}
        x^{+}={\;}&\boldsymbol{A}\left(\eta\right) x+ \boldsymbol{E} w, \\
        y={\;}&\boldsymbol{C} x, \\
    \end{aligned}
\end{equation}
with
\begin{subequations}
    \begin{align}
        \boldsymbol{A}(\eta) &= \operatorname{diag}\left(A(\eta^{u,l}_a), A(\eta^{u,l}_b), A(\eta^{u,l}_c)\right),\\
        % \boldsymbol{A}(\eta) &= \operatorname{diag}\left(A(\eta^u_a,\eta^l_a), A(\eta^u_b,\eta^l_b), A(\eta^u_c,\eta^l_c)\right),\\
        \boldsymbol{E}&= [\operatorname{diag}(E_a, E_b, E_c), E_z],\\
        \boldsymbol{C}&= \operatorname{diag}(C_a, C_b, C_c), 
    \end{align}
\end{subequations}
where, for all phases $m \in \{a,b,c\}$,
\begin{subequations}
    \begin{align*}
        A\left(\eta_m^{u,l}\right)&=\left[\begin{array}{cccc}
                                        K_1 & 0 & K_2 \eta_m^u & 0 \\
                                        0 & K_1 & 0 & K_2 \eta_m^l \\
                                        K_3 \eta_m^u & 0 & 1 & 0 \\
                                        0 & K_3 \eta_m^l & 0 & 1
                                        \end{array}\right],\\
        E_m &= \left[ \begin{array}{cc}
                         K_2 & 0   \\
                        -K_2 & 0 
                    \end{array}\right],  \\        
          C_m &= \left[\begin{array}{cccc}
                         1 & -1 & 0 & 0  \\
                         0.5 & 0.5 & 0 & 0 
                        \end{array}\right],    
    \end{align*}
\end{subequations}
and $E_z = \left[ \begin{array}{cc}
                         -K_2 \mathbf{1}_6 & \mathbf{0}_6   
                \end{array}\right].$
% \begin{equation}
%     E_z = \left[ \begin{array}{cc}
%                          -K_2 \mathbf{1}_6 & \mathbf{0}_6   
%                 \end{array}\right].
% \end{equation}
The constants $K_1 {=} {1- \frac{R_m T_s}{L_m}}$, $K_2 {=} \frac{T_s}{L_m}$, $K_3 {=} {-}\frac{N T_s}{C_{\gamma}}$ were obtained by discretizing \eqref{eq:DifferentialEquationUpper} with sampling time $T_s$, using the forward Euler method.

\textcolor{black}{The complements of the grid and MMC output voltages}, i.e., $v^{g'}_m$ and $v^{z'}$, allow to model the dynamics of the exogenous inputs as an autonomous system, i.e., 
\begin{equation}
\label{eq:exgenousSystemGlobal}
    w^+ = \boldsymbol{S} w,
\end{equation}
with $\boldsymbol{S} =  \operatorname{diag}(\sigma(\omega_1 T_s), \sigma(\omega_1 T_s), \sigma(\omega_1 T_s), \sigma(\omega_2 T_s)),$
where $\sigma(\cdot): \mathbb{R} \rightarrow \mathbb{R}^{2\times 2}$, such that
\begin{equation}
        \sigma({\omega T_s}){=}\left[\begin{array}{cc}
        \cos \left({\omega T_s}\right) & \sin \left({\omega T_s}\right) \\
        -\sin \left({\omega T_s}\right) & \cos \left({\omega T_s}\right)
\end{array}\right], 
% \end{align}
\end{equation}
 $T_s {\in} \mathbb{R}$ is the sampling time, $\omega_1  = 2 \pi f_1 \in \mathbb{R}$ is the grid angular frequency, and $\omega_2 = 2 \pi f_2 \in \mathbb{R}$ is the MMC output angular frequency.
% As a consequence of using \eqref{eq:exgenousSystemGlobal}, there is no need for PLL since all the variables can be directly or indirectly measured.
% As a consequence of \eqref{eq:exgenousSystemGlobal}, we can estimate the exogenous system dynamics using previous measurements. Hence, there is no need for PLL.

Albeit accurate, \eqref{eq:mmcBilinearModelGlobal} is \textcolor{black}{not commonly} used when designing conventional control methods for the MMC \cite{Du2018MMCBook, Sharifabadi2016MMCBook} due to the bilinear dynamics. 
A physics-informed transformation that simplifies \eqref{eq:mmcBilinearModelGlobal} is to assume the arm voltages as the control inputs, i.e, \textcolor{black}{disregard the capacitor voltage dynamics.}
Consequently, by defining a control input vector, i.e., $ u = [u_a, u_b, u_c]^\top$  with
\begin{equation}
    u_m = \left[\begin{array}{cc}u^u_m & u^l_m \end{array}\right], \; \forall m{\in}\{a,b,c\},
\end{equation} the total arm voltage dynamics are disregarded. This results in a reduced state vector, i.e., $\bar{x} =[\bar{x}_a, \bar{x}_b, \bar{x}_c]^\top$ with
\begin{equation}
    \bar{x}_m = \left[\begin{array}{cc}{\iota}^u_m &{\iota}^l_m\end{array}\right], \; \forall m{\in}\{a,b,c\},
\end{equation} and in a linear state-space model, i.e., 
 \begin{equation}
\label{eq:mmcLinearModelGlobal}
    \begin{aligned}
        \bar x^{+}={\;}&{\boldsymbol{\boldsymbol{\bar{A}}}} \bar x+ \boldsymbol{\bar B} u + \boldsymbol{E} w,\\
        y={\;}& {\boldsymbol{\Bar{C}}} \bar x, \\
    \end{aligned}
\end{equation}
with
\begin{subequations}
    \begin{align}
        {\boldsymbol{\boldsymbol{\bar{A}}}} &= \operatorname{diag}(\bar A_a, \bar A_b, \bar A_c),\\
        \boldsymbol{\bar B} &= \operatorname{diag}(\bar B_a, \bar B_b, \bar B_c),\\
        {\boldsymbol{\Bar{C}}} &= \operatorname{diag}(\bar C_a, \bar C_b, \bar C_c),
    \end{align}
\end{subequations}
where for all phases $m \in \{a,b,c\}$,
\begin{subequations}
\begin{align}
       \bar A_m &=\left[ \begin{array}{cc}
         K_1 & 0  \\
         0 &  K_1
    \end{array}\right], \;
        B_m = \left[ \begin{array}{cc}
         K_2 & 0  \\
         0 &  K_2
    \end{array} \right], \\ 
       \bar C_m &= \left[ \begin{array}{cccc}
         1 & -1   \\
         0.5 & 0.5 
    \end{array}\right].
\end{align} 
\end{subequations}

% ============================================================================================================================================
% ========================================================  MMC Control Problem  =============================================================
% ============================================================================================================================================

\section{MMC Control Problem Description}
\label{sec:mmcControlProblem}
\begin{figure}[htbp]
\centerline{\includegraphics[width=0.7\columnwidth]{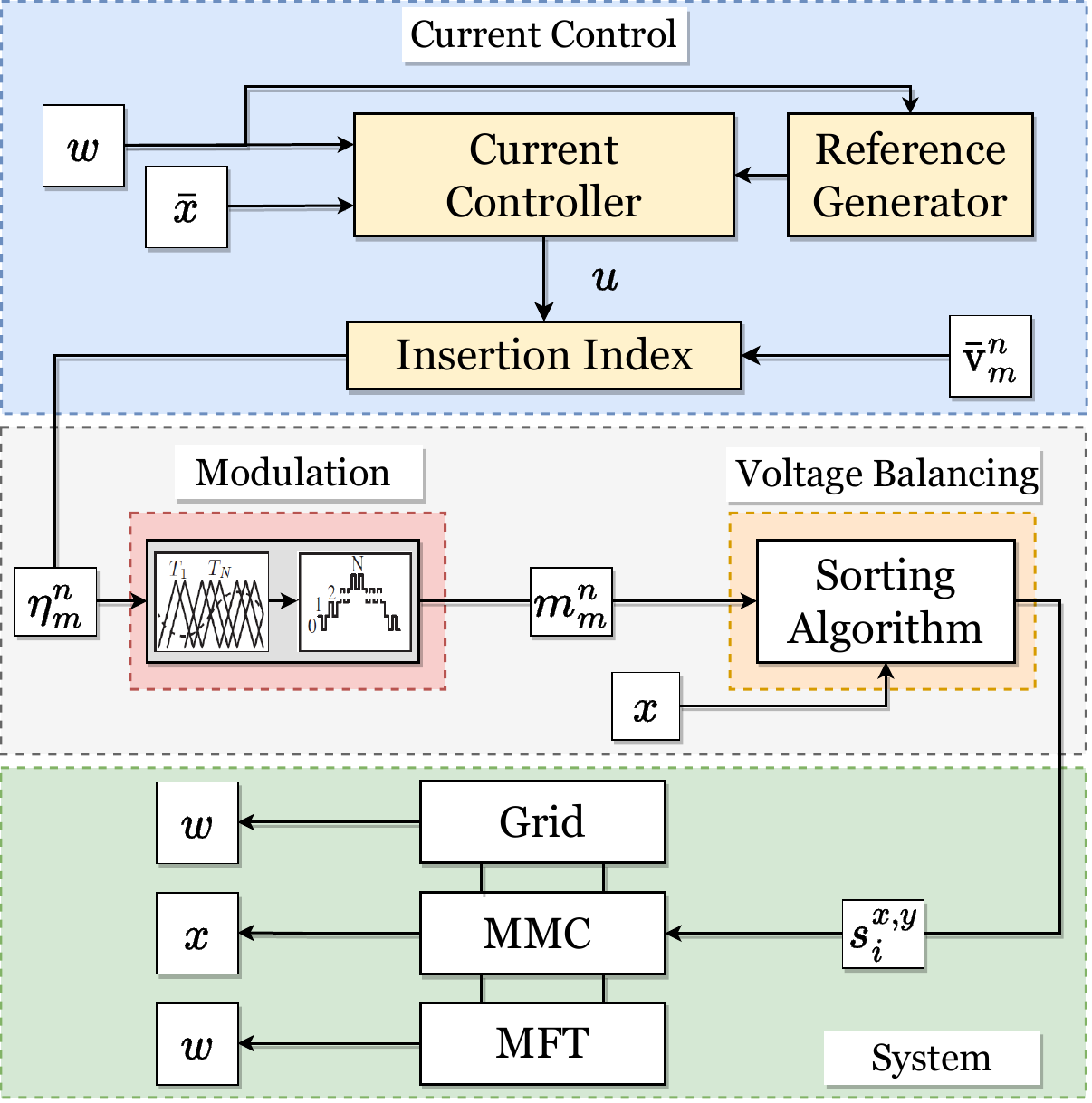}}
% This figure comes from the file: PlanPhD.drawio
% \centerline{\includesvg[width=\columnwidth]{Figures/2. Current Control Fig/CurrentControlPhDLandscape.svg}}
\caption{\textcolor{black}{Typical hierarchical control scheme of CCSs, as in \cite{Du2018MMCBook}, for MMCs connected to the grid and a medium frequency transformer (MFT).}}
\label{FigChap3:MMC_LOT_LM_CC_Vertical}
\end{figure}

Figure~\ref{FigChap3:MMC_LOT_LM_CC_Vertical} shows the most common hierarchical control scheme of MMCs \cite{Du2018MMCBook}.
This scheme has three stages: (i) current control, (ii) modulation, and (iii) voltage balancing, which are described in detail in \cite{Du2018MMCBook,Sharifabadi2016MMCBook}.
This paper focuses on the first stage, i.e., current control, which directly regulates the grid and the MMC output current as follows:
\begin{subequations}
\label{eq:currentControlProblem}
    \begin{align}
        \lim_{k \rightarrow \infty} & e^g_m :=\iota^g_m(k) - \iota^{g*}_m(k) = 0, \quad \forall m \in \{a,b,c\}, \\  
        \lim_{k \rightarrow \infty} &e^z_m := \iota^z_m(k) - \iota^{z*}_m(k) = 0, \quad \forall m \in \{a,b,c\},
    \end{align}
\end{subequations}
such that the desired per-phase power transfer is achieved, i.e.,   
\begin{equation}
    p^{g*}_m =v^{g}_m \iota^{g*}_m, \;\, p^{z*}_m = 2 v^{z} \iota^{z*}_m, \; \forall m \in \{a,b,c\}
\end{equation}
with 
\begin{subequations}
\label{eq:dynamicsReferences}
    \begin{align}
        \iota^{g*}_m &= \hat{I}_m^{g*} \cos(\omega_1 t + \phi_1), \;\, v^{g}_m= \hat{V}_m^{g} \cos(\omega_1 t),\\
        \iota^{z*}_m &= \hat{I}_m^{z*} \cos(\omega_2 t + \phi_2), \;\, \;\;  v^{z}= \hat{V}^{z} \cos(\omega_2 t ),
    \end{align}
\end{subequations}
where $p^{g*}_m$ and $p^{z*}_m$ are the desired instantaneous power of the grid and transformer corresponding to the phase $m$; $\hat{I}_m^{g*}$, $\hat{I}_m^{z*}$, $\hat{V}_m^{g}$ and $\hat{V}^{z}$ the peak-to-peak value of each corresponding variables; and $\phi_1 $ and $\phi_2$ are the phase angle of the currents.

\textcolor{black}{To design the references $\iota^{g*}_m$ and $\iota^{z*}_m$, we use the common- and differential-mode decomposition presented in \cite{Pereira2021ExampleACAC}, i.e.,} 
\begin{subequations}
\label{eq:defCommonDiffMode}
    \begin{align}
        \iota^{\Delta}_m &=  \frac{\iota^u_m -  \iota^l_m}{2}, \, u^{\Delta}_m =  \frac{u^u_m -  u^l_m}{2}, \,\mathrm{v}^{\Delta}_m =  \frac{\mathrm{v}^u_m -  \mathrm{v}^l_m}{2}, \\
        \iota^{\Sigma}_m &= \frac{\iota^u_m +  \iota^l_m}{2}, \, u^{\Sigma}_m =  \frac{u^u_m +  u^l_m}{2}, \, \mathrm{v}^{\Sigma}_m = \frac{\mathrm{v}^u_m +  \mathrm{v}^l_m}{2},
    \end{align}
\end{subequations}
\textcolor{black}{where the superscripts $^{\Delta}$ and $^{\Sigma}$ label the differential and common-mode components, respectively. }
\textcolor{black}{This decomposition decouples the dynamics of the grid and the MMC output currents, since $\iota^{g}_m =  2 \iota^{\Delta}_m$ and $\iota^{z}_m =  \iota^{\Sigma}_m$ \cite{Pereira2021ExampleACAC,Fuchs2020LQR}.}
\textcolor{black}{Using \eqref{eq:defCommonDiffMode} and the MMC dynamics in \eqref{eq:mmcLinearModelGlobal}, we propose a quadratic optimization problem to compute the optimal references.}
\textcolor{black}{In this case, given  $\iota^{\Delta}_m$, we can compute $\iota^{\Sigma}_m$, such that}
\begin{subequations}
\label{eq:minimizationProblemReferences}
\begin{align}
    \label{eq:minimizationProblemReferences1}
    \min _{I_m^{\Sigma}} & \left\| \sum_{k=1}^{N_t(f_1)} \frac{\left(v_m^g(k) \iota_m^{\Delta}(k)\right)}{N_t(f_1)} - \sum_{k=1}^{N_t(f_2)}\frac{\left(v_m^z(k) \iota_m^{\Sigma}(k)\right)}{N_t(f_2)}\right\|, \\
    \label{eq:minimizationProblemReferences2}
\text { s.t. } & \sum_{k=1}^{N_t(f_1)}  \frac{\left(u_m^{\Delta}(k) \iota_m^{\Delta}(k)\right)}{N_t(f_1)}+\sum_{k=1}^{N_t(f_2)}\frac{\left(u_m^{\Sigma}(k) \iota_m^{\Sigma}(k)\right)}{N_t(f_2)} =0,
\end{align}
\end{subequations}
\textcolor{black}{where $N_t(f) =  1/f T_s \in  \mathbb{N}$ is the number of samples in a period, i.e., $T = 1/f$. }
\textcolor{black}{Using the power equations in \cite{IEEE_STD_2010}, we calculate $\iota^{\Delta*}_m$  and assume that $2\iota^{\Delta*}_m =  \iota^{g*}_m$.}
\textcolor{black}{Since \eqref{eq:minimizationProblemReferences} considers the dynamics of \eqref{eq:mmcLinearModelGlobal}, we compute an optimal $\iota_m^{\Sigma}$, i.e., $\iota_m^{\Sigma*}$, which minimizes the power difference between grid and the MMC output while keeping the average of the arm power equal to zero.} 
\textcolor{black}{As explained in \cite{Pereira2021ExampleACAC}, imposing \eqref{eq:minimizationProblemReferences2} ensures  stable dynamics of the total arm voltage, such that, for all phases $m \in \{a,b,c\}$,}
\begin{equation}
\label{eq:sotredVoltagetControlProblem}
% \begin{aligned}
    \sum^{N_T}_{j=1}\mathrm{v}^{\Delta}_m(k+j) =0, \;\text{and}\,
    \sum^{N_T}_{j=1} \mathrm{v}^{\Sigma}_m(k+j) = \hat{\mathrm{V}}^g_m + \hat{\mathrm{V}}^z.
% \end{aligned}     
\end{equation}
\textcolor{black}{Hence, by tracking these references, the current controller generates arm voltages that guarantee safe operation of the MMC.}
From \eqref{eq:dynamicsReferences}, we infer the linear relationship between the exogenous inputs and the reference vector for the current controller, i.e., $r = [r_a, r_b, r_c]^\top$ with 
\begin{equation}
    r_m = [\iota^{g*}_m \;\, \iota^{z*}_m], \quad \forall m \in \{a,b,c\},
\end{equation}
such that
\begin{equation}
    r = \boldsymbol{O} w
\end{equation}
with $\boldsymbol{O} =\left[\operatorname{diag}(O_a,O_b,O_c), [O_z; O_z;O_z]\right]$, where
\begin{subequations}
    \begin{align}
        O_m&{=}\left[\begin{array}{cc} o_1 & o_2 \\ 0 & 0 \end{array}\right], \quad \forall m \in \{a,b,c\},\\
        O_z&{=}\left[\begin{array}{cc}  0 & 0 \\ o_3 & o_4 \end{array}\right].
    \end{align}
\end{subequations}
The coefficients of $\boldsymbol{O}$ are $o_{1} = K_4 \cos{ \left(\phi_1\right)}$, $o_{2} = -K_4 \sin{ \left(\phi_1\right)}$, $o_{3}=K_5 \cos{ \left(\phi_2\right)}$ and $o_{4} = -K_5 \sin{ \left(\phi_2\right)}$ with $K_4 {=} 2 \hat{I}_m^{\Delta*}/ \hat{V}_m^{{g}}$ and $K_5 {=} \hat{I}_m^{\Sigma*}/ \hat{V}_m^{{z}}$, as in  \cite{IEEE_STD_2010}.

% The current controller only calculates the arm voltage ($u^n_m$), so the control problem expressed in \eqref{eq:currentControlProblem} is solved.
As in \eqref{eq:mmcLinearModelGlobal}, the control input of the current controller is the arm voltage ($u^n_m$).
Thus, to transform the arm voltage into an insertion index, we use the formulation from \cite{Harnefors2015MMC17}, i.e.,
% In the next stage of the hierarchical scheme, this insertion index is transformed into the switching signal that ensures stable operation of the MMC.
% We calculate the insertion index as in \cite{Harnefors2015MMC17}, i.e.,
    \begin{equation}
    \label{eqChap3:InsertionIndexTranslationDirect}
        \eta^n_m =   \operatorname{sat}\left(\frac{u^n_m}{ \hat{\mathrm{V}}^g_m + \hat{\mathrm{V}}^z_m}\right), \forall n \in \{u,l\}, m \in \{a,b,c\}
    \end{equation}
    with $\operatorname{sat}(\cdot): \mathbb{R} \rightarrow \mathbb{R}_{[-1,1]}$, i.e., 
    \begin{equation}
    \label{eqChap3:StaturationFormulation}
    \operatorname{sat}(\xi):=\left\{\begin{array}{cc}
    1 & \text { if } \xi>1 \\
    \xi & \text { if }|\xi| \leq 1 \\
    -1 & \text { if } \xi<-1
    \end{array}\right. .
    \end{equation}
\textcolor{black}{Using \eqref{eq:minimizationProblemReferences} will ensure the stabilising properties of \eqref{eqChap3:InsertionIndexTranslationDirect} as we will show in the next sections.}    

\section{MIMO Current Control for MMC}
\label{sec:mmcCurrentController}

In this section, we present a framework to design a multiple-input, multiple-output (MIMO) controller to track the currents of the MMC as in \eqref{eq:currentControlProblem}, while stabilising the total arm voltage as in \eqref{eq:sotredVoltagetControlProblem}.
In this case, we aim to reach zero tracking error ($e(k)$), such that
\begin{equation}
\label{eq:errorGeneralDef1}
    \lim_{k \rightarrow \infty} e(k):= y(k) - r(k) = 0,
\end{equation}
while meeting performance and safety constraints, i.e., 
\begin{equation}
    x(k) \in \mathcal{X} \subseteq \mathbb{R}^{12}, \; u(k) \in \mathcal{U}\subseteq \mathbb{R}^6 \;\, \forall k \in \mathbb{N}.
\end{equation}
Working in the $abc$-reference framework implies tracking time-varying references. Hence, the constraints are easily formulated as the error between the measurements and the ideal dynamics.
In this paper, we define the ideal dynamics of the arm currents as the steady-state corresponding to reaching zero tracking error as in \eqref{eq:errorGeneralDef1}, i.e., 
\begin{equation}
\label{eq:idealDynamics}
    {\bar{x}^{ss^+}} =  \boldsymbol{\bar{A}} \bar x^{ss} + \bar{\boldsymbol{B}} u^{ss} + \boldsymbol{E} w,
\end{equation}
where $\bar x^{ss}$ is the reference steady-state, such that $r = \boldsymbol{\bar{C}} \bar x^{ss}$ and $u^{ss}$ is the steady-state of the control input.
Then, let us define the state tracking error as
\begin{equation}
\label{eq:stateErrorDefnition}
    e_x(k) := \bar x(k)- \bar x^{ss}(k), 
\end{equation} 
and control input tracking error as  
\begin{equation}
\label{eq:scontrolInptErrorDefnition}
    e_u(k) :=  u(k)- u^{ss}(k).
\end{equation}
Then, the controller must meet the error constraints, i.e., 
\begin{equation}
    % e_x(k) \in \mathbb{X} \; \text{and} \; e_u(k) \in \mathbb{U}, \, \text{and} \, e(k)\in \bar C \mathbb{X}   \;\,\forall k \in \mathbb{N}  
    e_x(k) \in \mathbb{X} \; \text{and} \; e_u(k) \in \mathbb{U},  \;\,\forall k \in \mathbb{N},
\end{equation}
with
\begin{subequations}
    \label{eq:constraintDefinitionLemma}
        \begin{align}
            \mathbb{X} &:= \{e_x\in\mathbb{R}^{6}: g_{t_x}^\top e_x \leq 1, \forall t_x \in \{1,{...},s\}\},\\
            \mathbb{U} &:= \{ e_u\in\mathbb{R}^{6}: h_{t_u}^\top  e_u \leq 1, \forall t_u \in \{1,{...},l\}\}.
        \end{align}
\end{subequations}
% \begin{subequations}
%     \label{eq:constraintDefinitionLemma}
%         \begin{align}
%             \mathbb{X} &:= \{\bar x- \bar{x}_{ss}\in\mathbb{R}^{6}: g_{t_x}^\top \bar x \leq 1, \forall t_x \in \{1,{...},s_i\}\},\\
%             \mathbb{U} &:= \{u - u^{ss}\in\mathbb{R}^{6}: h_{t_u}^\top u \leq 1, \forall t_u \in \{1,{...},l_i\}\}.
%         \end{align}
% \end{subequations}
In this case, $e_x(k) \in \mathbb{X} \implies  x \in \mathcal{X}$ and $e_u(k) \in \mathbb{U} \implies u \in \mathcal{U}$, i.e., we define a bound for tracking the MMC arm currents such that the total arm voltage is also kept bounded within a certain operation region.
% where $\bar x^{ss}$ is the steady-state corresponding to the reference, i.e., $r = \bar C \bar x^{ss}$ and $u^{ss}$ is the control input that guarantees to reach that steady-state.
% Consequently, we can also obtain a bounded tracking error, i.e.,  
% \begin{equation}
% \label{eq:errorGeneralDef1}
%     e(k):= y(k)-r(k) \in \bar C \mathbb{X}.
% \end{equation}

To design the proposed controller, we consider a compact representation of the MMC, grid and transformer described by \eqref{eq:mmcLinearModelGlobal} and \eqref{eq:exgenousSystemGlobal}, i.e.,
\begin{equation}
\label{eq:compactLinerMMCModel}
\bar{\Sigma}:\left\{\begin{array}{l}
\,\bar{x}^{+}=\boldsymbol{\bar{A}} \bar{x}+\boldsymbol{\bar B} u+\boldsymbol{E} w, \\
\;\;\,\,y=\boldsymbol{\bar{C}} \bar{x}, \\
\hline
w^{+}=\boldsymbol{S} w, \\
\;\;\,\,r=\boldsymbol{O} w,
\end{array}.\right.
\end{equation}
In \cite{Francis1976LOT}, the authors discussed an unconstrained solution for \eqref{eq:errorGeneralDef1} when linear dynamics as in \eqref{eq:mmcLinearModelGlobal} are considered, i.e., the linear output tracking problem (LOTP). From \cite{Francis1976LOT}, it is known that there exists a solution to the LOTP if the following assumption holds.
% \cite[Theorem 1]{Francis1976LOT}, it is known that there exist a family of static feedback gain controllers as 
% \begin{equation}
% \label{eq:definitionMIMOController}
%     u = K_x \bar x + K_w w,
% \end{equation}
% that can ensure asymptotic zero tracking error, i.e., \eqref{eq:errorGeneralDef1}, in closed-loop with \eqref{eq:compactLinerMMCModel}
% % \begin{equation}
% % \label{eq:asymptoticError}
% %     \lim_{k \rightarrow \infty} e(k) = 0,
% % \end{equation}
% if the following assumption holds.
\begin{assumption}[Sylvester's equation for the LOTP \cite{Francis1976LOT}]
\label{amp:theAssumption}
    There exist matrices $\Pi$ and $\Gamma$ such that 
    \begin{subequations}
    \label{eqChap3:SylvesterEqFromWonham}
        \begin{align}
        & \Pi S=A \Pi+B \Gamma+E, \\
        & C \Pi=O,
    \end{align}
    \end{subequations}
    where $A$, $B$, $C$, $E$, are the matrices of the state space model of a linear system, and $S$ and $O$ are the matrices describing an autonomous exogenous system, as in \eqref{eq:compactLinerMMCModel}.
\end{assumption}   

% for any linear system as \eqref{eq:compactLinerMMCModel}. 
% Using this framework, we propose to find a feedback gain that meets our criteria using the following lemma:
\begin{theorem}[Controller Synthesis]
\label{lmm:ControllerSynthesis}
    Suppose that Assumption~\ref{amp:theAssumption} holds for system \eqref{eq:compactLinerMMCModel}.
    Consider the following linear matrix inequalities:
    \begin{subequations}
    \label{eq:LMISMIMO}
        \begin{align}
            \left[\begin{array}{cc}
            Z & (\boldsymbol{\bar{A}} Z+\boldsymbol{\bar B} Y)^{\top} \\
            (\boldsymbol{\bar{A}} Z+\boldsymbol{\bar B} Y) & Z
        \end{array}\right] &\succ 0,\\
        \forall t_x \in\left\{1, \ldots, s\right\}, \quad\left[\begin{array}{cc}
            Z & \left(Z g_{t_x}\right)^{\top} \\
            \left(Z g_{t_x}\right) & 1
        \end{array}\right] &\succcurlyeq 0,\\
        \forall t_u \in\left\{1, \ldots, l\right\}, \quad\left[\begin{array}{cc}
            Z & \left(Y h_{t_u}\right)^{\top} \\
            \left(Y h_{t_u}\right) & 1
        \end{array}\right] &\succcurlyeq 0,    
        \end{align}
    \end{subequations}
    where the pair ($\boldsymbol{\bar{A}},\boldsymbol{\bar B}$) comes from \eqref{eq:mmcLinearModelGlobal}, $g_{t_x}$ and $h_{t_u}$ are vectors corresponding to the error constraints from $\mathbb{X}$ and $\mathbb{U}$, respectively.
    If there exist $Y$ and $Z$ such that \eqref{eq:LMISMIMO} holds, then the closed-loop system comprised by \eqref{eq:compactLinerMMCModel} and the static feedback gain controller, i.e., 
    \begin{equation}
    \label{eq:definitionMIMOController}
        u = K_x \bar x + K_w w,
    \end{equation}
    where 
    \begin{equation}
        K_x =  YZ^{-1}\; \text{and} \;K_w=\Gamma-K_x \Pi,
    \end{equation} is asymptotically stable with a domain of attraction, i.e., 
    \begin{equation}
        \mathbb{S}_x=\left\{e_x \in \mathbb{R}^{6}: e_x^{\top} P e_x \leq 1\right\} \subseteq \mathbb{X},
    \end{equation}
    where $P=Z^{-1}$, and the level set of feasible control inputs, i.e., $\mathbb{S}_u = K_x \mathbb{S}_x \subseteq \mathbb{U}$.    
\end{theorem}

\begin{IEEEproof}[\textit{Proof}]
From \cite{Francis1976LOT}, it is known that given the pair ($\Pi, \Gamma$), the steady-state dynamics are defined as 
\begin{equation}
\label{eq:ssDefinitionProof}
    \bar x^{ss} = \Pi w, \; \text{and} \; u^{ss} =  \Gamma w
\end{equation}
yields the ideal dynamics as in \eqref{eq:idealDynamics}, where $e(k) = \bar C \bar x^{ss}(k) - r(k) = 0$, for all $k \in \mathbb{N}$.
Using \eqref{eq:stateErrorDefnition}, \eqref{eq:scontrolInptErrorDefnition}, \eqref{eq:compactLinerMMCModel} and \eqref{eq:ssDefinitionProof}, we derive the state tracking error dynamics and the control input error as follows
\begin{subequations}
\label{eq:errorDefinitionProof}
    \begin{align}
        e^+_x &= (\boldsymbol{\bar{A}}  + \boldsymbol{\bar B} K_x) e_x,\\
        e_u &= K_x e_x.
    \end{align}
\end{subequations}
The full derivation of \eqref{eq:errorDefinitionProof} can be founded in Appendix\ref{appx:DerivationErrorDynamics}.
Note that the LMIs \eqref{eq:LMISMIMO} can be rewritten as:
    \begin{subequations}
\label{eq:original inequalities}
    \begin{align}
        (\boldsymbol{\bar{A}}  + \boldsymbol{\bar B} K_x)^\top P (\boldsymbol{\bar{A}}  + \boldsymbol{\bar B} K_x) \prec  P,\label{eq:oriLMI_1}\\
        \forall t_x {\in }\{1,2,{...},s_i\},\quad \sqrt{g_{t_x}^\top P^{-1} g_{t_x}} \leq 1,\label{eq:oriLMI_3}\\
        \forall t_u {\in}  \{1,2,{...},l_i\},\quad \sqrt{h_{t_u}^\top K_x P^{-1} K_x^\top h_{t_u}} \leq 1,\label{eq:oriLMI_4}
    \end{align} 
\end{subequations}
where \eqref{eq:oriLMI_1}{-}\eqref{eq:oriLMI_4} are obtained by applying Schur complement, respectively. If \eqref{eq:original inequalities} is feasible for a given $\boldsymbol{\bar{A}} $, $\boldsymbol{\bar B}$,  $g_{t_x}$ and $h_{t_u}$, then
\romannumeral 1) $\rho(\boldsymbol{\bar{A}}  + \boldsymbol{\bar B} K_x) < 1$ that implies that $\lim_{k \rightarrow \infty} e_x(k)  =0$ and $\lim_{k \rightarrow \infty} e_u(k)  =0$, i.e., the state tracking error is asymptotically stable, and consequently, the LOTP is solved, \romannumeral 2) $(\boldsymbol{\bar{A}}  + \boldsymbol{\bar B}K_x)\mathbb{S}_x \subseteq \mathbb{S}_x$, i.e., $\mathbb{S}_x$ is a positively invariant set, \romannumeral 3) $\mathbb{S}_x \subseteq \mathbb{X}$, i.e., the domain of attraction is constraint admissible \romannumeral 4) $K_x\mathbb{S}_x \subseteq \mathbb{U}$, i.e., the resulting level set is constraint admissible too.
\end{IEEEproof}

\begin{remark}
    For Theorem~\ref{lmm:ControllerSynthesis}, we assume that $K_x$ is not given. 
    However, to ensure certain error dynamics, $K_x$ can be pre-calculated and then, solve the LMIs \eqref{eq:LMISMIMO}.
    The main difference with Theorem~\ref{lmm:ControllerSynthesis} is that the size of $\mathbb{S}_x$ can be affected by $K_x$.
    This approach can boost MMC's performance, but it may reduce the domain of attraction.
\end{remark}

\begin{figure}[htbp]
\centerline{\includegraphics[width=\columnwidth]{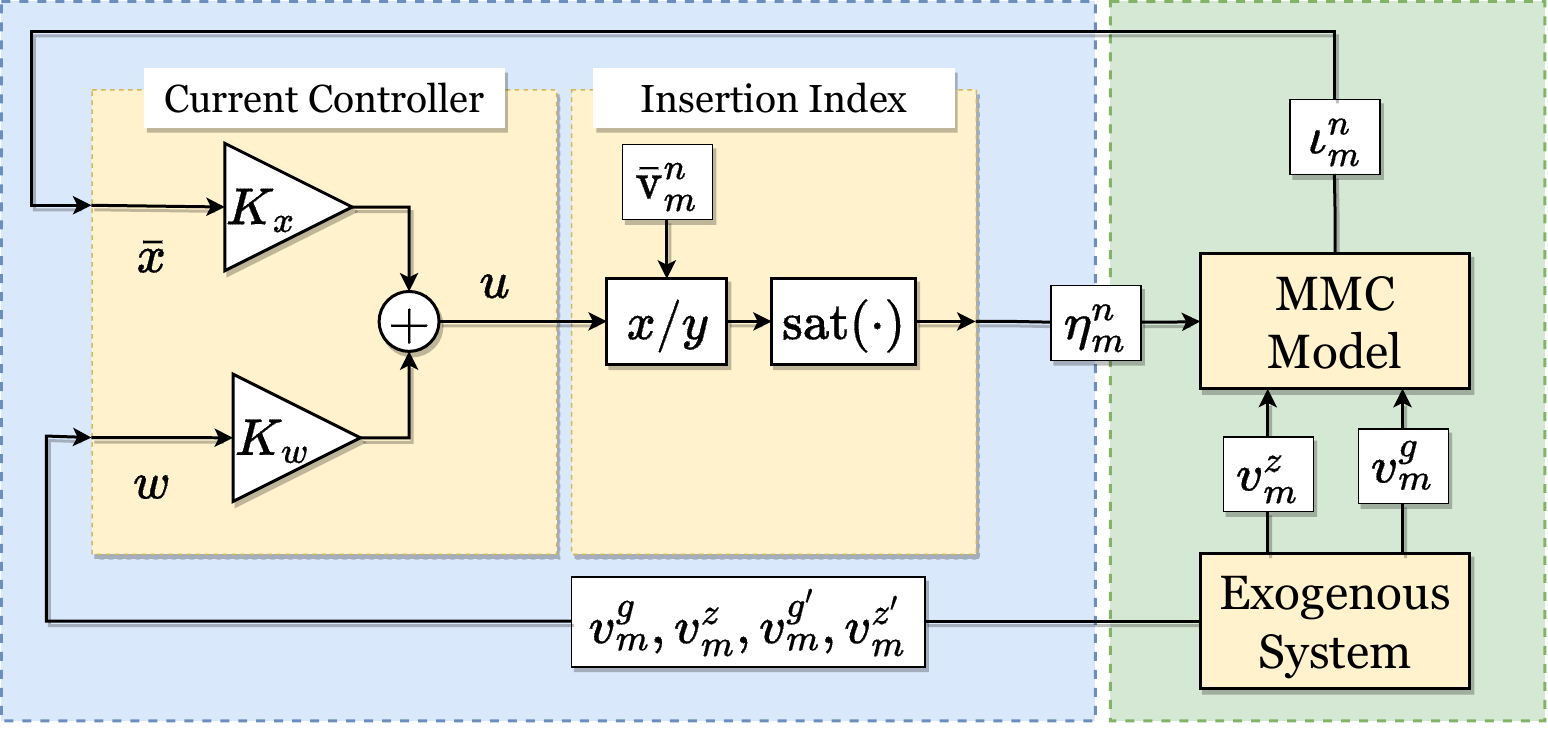}}
% This figure comes from the file: PlanPhD.drawio
% \centerline{\includesvg[width=\columnwidth]{Figures/2. Current Control Fig/CurrentControlPhDLandscape.svg}}
\caption{Block diagram of the proposed MMC's Current Control}
\label{FigChap3:MMC_LOT_LM_CC}
\end{figure}

\begin{remark}
    \textcolor{black}{Employing the developed controller results in a structure of the current control as shown in Figure~\ref{FigChap3:MMC_LOT_LM_CC}.}
    \textcolor{black}{This controller works in the $abc$-reference framework and only needs measurements of the grid and the MMC output voltages to align to their phases.}
    \textcolor{black}{In this case, we are not limited by the shape of the waveform as long as it can be modelled as a linear autonomous system as in \eqref{eq:exgenousSystemGlobal}.}
    \textcolor{black}{Moreover, the computational load is similar to the existent CCSs.}
    % \textcolor{black}{Hence, this controller}
\end{remark}

% This control scheme is simpler than the CCSs, yet performance and computational complexity are not compromised. 
% Moreover, this controller works in the $abc$-reference framework; hence, it does not need the use of PLL.

% ============================================================================================================================================
% ========================================================  Stability Guarantees  ============================================================
% ============================================================================================================================================
\subsection{Stability Guarantees}
\label{subsec:stabilityGuarantees}
Theorem~\ref{lmm:ControllerSynthesis} proves the convergence and constraint satisfaction of the closed-loop system \eqref{eq:compactLinerMMCModel}-\eqref{eq:definitionMIMOController}. 
However, \eqref{eq:compactLinerMMCModel} disregards the use of the insertion index and the dynamics of the total arm voltage ($\mathrm{v}^n_m$).
Hence, we still need to ensure MMC stability, i.e., safe operation, when the controller \eqref{eq:definitionMIMOController} is used.

Let us consider an equivalent representation of \eqref{eq:compactLinerMMCModel}, i.e., a bilinear model, such that
\begin{equation}
\label{eq:compactAugementedLinearMMC}
\tilde{\Sigma}:\left\{\begin{array}{l}
\,\tilde{x}^{+}=\boldsymbol{\tilde{A}}\left(\eta\right) \tilde{x}+\boldsymbol{E} w, \\
 \;\;\,\,y=\boldsymbol{C} \tilde{x}, \\
\hline
w^{+}=\boldsymbol{S} w, \\
\;\;\,\,r= \boldsymbol{O} w,
\end{array},\right.
\end{equation}
with $ \boldsymbol{\tilde{A}}(\eta) = \text{diag}\left(\tilde A(\eta^{u,l}_a), \tilde A(\eta^{u,l}_b), \tilde A(\eta^{u,l}_c)\right),$ and $\tilde x =[\tilde x_a, \tilde x_b, \tilde x_c]^\top$, where for all phases $m \in \{a,b,c\}$,
\begin{subequations}
\label{eq:compactAugementedLinearMatrixMMC}
    \begin{align}
    \tilde{A}\left(\eta_m^{u,l}\right)&=\left[\begin{array}{cccc}
                                                K_1 & 0 & K_2 \eta_m^u & 0 \\
                                                0 & K_1 & 0 & K_2 \eta_m^l \\
                                                0 & 0 & 1 & 0 \\
                                                0 & 0 & 0 & 1
                                                \end{array}\right], \; \text{and}\\
        \tilde{x}_m &=\left[{\iota}_m^u \, {\iota}_m^l \, \mathrm{v}_m^u\, \mathrm{v}_m^l\right].
    \end{align}
\end{subequations}
In \eqref{eq:compactAugementedLinearMMC}, we incorporate the bilinear dynamics caused by $\eta$. However, the total arm voltage has decoupled constant dynamics, i.e.,
\begin{equation}
    \mathrm{v}_m^n(k) = \mathrm{v}_m^n(0), \quad \forall k \in \mathbb{N}_+. 
\end{equation}
Even when \eqref{eq:compactAugementedLinearMMC} is not physically accurate, it ensures that  
\begin{equation}
    \boldsymbol{C} \tilde x = \boldsymbol{\bar{C}} \bar x \quad \forall k \in \mathbb{N}.
\end{equation}
\textcolor{black}{Considering the references resulting from \eqref{eq:minimizationProblemReferences} and $\mathrm{v}_m^n(0) = \hat{\mathrm{V}}^g_m + \hat{\mathrm{V}}^z_m $ implies that} the closed-loop system \eqref{eq:definitionMIMOController}-\eqref{eq:compactAugementedLinearMMC} tends to the ideal dynamics of the MMC presented in \cite{Sharifabadi2016MMCBook, Pereira2021ExampleACAC}; i.e., $ \lim_{k\rightarrow \infty}\tilde x_m (k) =  x^*_m (k)$, for all $k \in \mathbb{N}$ with
\begin{equation}
     x^*_m (k)  = \left[{\iota}_m^{u*}\, {\iota}_m^{l*}\, \mathrm{v}_m^{u}\, \mathrm{v}_m^{l}\right], \forall m \in \{a,b,c\}
\end{equation}
where ${\iota}_m^{u*} = {\iota}^{z*}_m +0.5 {\iota}^{g*}_m$, ${\iota}_m^{l*} = {\iota}^{z*}_m -0.5 {\iota}^{g*}_m$.
So, the deviation between the MMC dynamics as \eqref{eq:mmcBilinearModelGlobal}  and the closed-loop system \eqref{eq:definitionMIMOController}-\eqref{eq:compactAugementedLinearMMC} determines the MMC stability.

\begin{definition}[MMC Safe Operation]
\label{pbm:MMCStability}
Let $\mathcal{X} \subseteq \mathbb{R}^{12}$ be the constraint admissible set, i.e., safe operation region, and let $\varepsilon(k) \in \mathbb{R}^{12}$ be the MMC deviation from the ideal dynamics, i.e.,
    \begin{equation}
    \label{eq:errorLinNonLinMMCDef}
        \varepsilon(k) := x(k) - \tilde x(k).
    \end{equation}
    % Let us define $\mathcal{E}:={ \{ \varepsilon\in\mathbb{R}^{12}: {\varepsilon}^{\top} Q  \varepsilon \leq 1\}}$, i.e., a safe set, such that $\varepsilon \in \mathcal{E} \implies x \in  \mathcal{X}$.
    The MMC dynamics represented by \eqref{eq:mmcBilinearModelGlobal} are stable if for any $\gamma >0$  there exist $\delta =  \delta(\gamma)>0$ and $\| \varepsilon(0) \| \leq  \delta$ 
    such that
    \begin{equation}
    \label{eq:errorLinNonLinMMCDef2223}
        \| \varepsilon(0) \| \leq  \delta \implies  \|\varepsilon(k)\|  \leq \gamma, \forall k \in \mathbb{N}_+,
    \end{equation}
    where $x(0) = \tilde x(0) \in  \mathcal{X}$ are the initial states.
\end{definition}

% In this case, we focus on proving stability instead of local asymptotic stability because the ripple of the total arm voltage is an inherent property of the MMC, then $x(k) \nrightarrow \tilde x^*(k)$.

Based on \eqref{eq:mmcBilinearModelGlobal} and \eqref{eq:compactAugementedLinearMMC}, the MMC deviation dynamics, i.e., $\varepsilon^+ = x^+ - \tilde{x}^+$, yield 
\begin{equation}
\label{eq:dynamicsDef1}
    \varepsilon^{+}=\boldsymbol{A}\left(\eta\right) x -\boldsymbol{\tilde{A}}\left(\eta\right) \tilde{x}. 
\end{equation}
By adding and subtracting $\boldsymbol{{A}}\left(\eta\right) \tilde x$, \eqref{eq:dynamicsDef1}  results in
    \begin{equation}
    \label{eq:StabilityProof76}
    \varepsilon^{+}=\boldsymbol{A}\left(\eta\right) \varepsilon +\boldsymbol{\tilde{B}}(\eta)  \tilde{x},
    \end{equation}
where $\boldsymbol{\tilde{B}}(\eta) = \boldsymbol{A}\left(\eta\right)-\boldsymbol{\tilde{A}}\left(\eta\right)$.
% In \eqref{eq:StabilityProof76}, the deviations of the MMC represent a non-autonomous system.
% Note that \eqref{eq:StabilityProof76} belongs also to the autonomous system, i.e., 
%     \begin{equation}
%     \label{eq:StabilityProof3}
%     \left[\begin{array}{c}
%     \varepsilon \\
%     \tilde{x} \\
%     w
%     \end{array}\right]^{+}=\left[\begin{array}{ccc}
%     \boldsymbol{A}\left(\eta\right) & \boldsymbol{\tilde{B}}(\eta) & 0 \\
%     0 & \boldsymbol{\tilde{A}}\left(\eta\right) & \boldsymbol{E} \\
%     0 & 0 & \boldsymbol{S}
%     \end{array}\right]\left[\begin{array}{c}
%     \varepsilon \\
%     \tilde{x} \\
%     w
%     \end{array}\right] .
%     \end{equation}
Note that \eqref{eq:StabilityProof76} is a linear parameter-varying (LPV) representation, where $\eta \in \Xi \subseteq \mathbb{R}^6_{[-1,1]}$ is the scheduling-variable, with $\Xi$ being a closed polyhedron that contains the origin in its interior.
Hence, using the LPV framework, we can analyse the stability of \eqref{eq:StabilityProof76}, which implies the safe operation of the MMC, as explained in Definition~\ref{pbm:MMCStability}.

    % \begin{equation}
    % \label{eq:StabilityProof2}
    %     \varepsilon^+=\boldsymbol{A}\left(\eta\right) x-\boldsymbol{\tilde{A}}\left(\eta\right) \tilde{x}.
    % \end{equation}    
    
\begin{theorem}[MMC Stability]
\label{lmm:mmcStability}
Consider the system \eqref{eq:StabilityProof76} with the scheduling-varible $\eta \in \Xi \subseteq \mathbb{R}^6_{[-1,1]}$. Let us define a Lyapunov candidate function $\text{V}(\varepsilon)$ , i.e., $\text{V}(\varepsilon):=\varepsilon^T \boldsymbol{Q} \varepsilon$. 
% a subset of the constraints of the insertion indexes per arm, i.e., $\Xi =\{\xi \in \mathbb{R}^2  : -\mathbf{1}_{2} \leq \xi \leq \mathbf{1}_{2}\}$, with a convex hull representation as
% \begin{equation}
%     \Xi :=\text{conv}\{\xi^1, ...,\xi^p\},\; 
% \end{equation} 
% where $p \in \mathbb{N}_+$, $\xi^i$  are linearly independent vectors representing the $i$th vertex of the polytope. 
If there exists a matrix $\boldsymbol{Q}\succ 0$ such that following inequality  
\begin{equation}
\label{eq:inequalityStabilityLMim}
    \left[\begin{array}{cc}
        \boldsymbol{Q} &  \boldsymbol{Q}\boldsymbol{A}(\xi^i)\\
        \boldsymbol{A}(\xi^i)^\top \boldsymbol{Q} & \boldsymbol{Q}
    \end{array}\right] \succ 0, 
\end{equation}
is satisfied for all vertices $\xi^i$ of $\Xi$, then \eqref{eq:errorLinNonLinMMCDef2223} is satisfied, i.e., the MMC operation is stable, as in Definition~\ref{pbm:MMCStability}.
\end{theorem}

% \begin{theorem}[MMC Stability]
% \label{lmm:mmcStability}
% Consider the system \eqref{eq:StabilityProof76}, the insertion indexes as in \eqref{eqChap3:InsertionIndexTranslationDirect}, i.e., $\eta \in \mathbb{R}^6_{[-1,1]}$. Let us define a Lyapunov candidate function $\text{V}(\varepsilon)$ , i.e., $\text{V}(\varepsilon):=\varepsilon^T P \varepsilon$, a subset of the constraints of the insertion indexes per arm, i.e., $\Xi =\{\xi \in \mathbb{R}^2  : -\mathbf{1}_{2} \leq \xi \leq \mathbf{1}_{2}\}$, with a convex hull representation as
% \begin{equation}
%     \Xi :=\text{conv}\{\xi^1, ...,\xi^p\},\; 
% \end{equation} 
% where $p \in \mathbb{N}_+$, $\xi^i$  are linearly independent vectors representing the $i$th vertex of the polytope. 
% If there exists a matrix $P\succ 0$ such that following inequality  
% \begin{equation}
% \label{eq:inequalityStabilityLMim}
%     \left[\begin{array}{cc}
%         P &  PA(\xi^i)\\
%         A(\xi^i)^\top P & P
%     \end{array}\right] \succ 0, 
% \end{equation}
% is satisfied for all vertices $\xi^i$ of $\Xi$, then \eqref{eq:errorLinNonLinMMCDef2223} is satisfied, i.e., the MMC operation is stable, as in Definition~\ref{pbm:MMCStability}.
% \end{theorem}

\begin{IEEEproof}[\textit{Proof}]
Consider \eqref{eq:StabilityProof76}, i.e., a linear parameter-varying (LPV) system, as 
    \begin{equation}
    \varepsilon^{+}=\boldsymbol{A}\left(\eta\right) \varepsilon +\boldsymbol{\tilde{B}}(\eta)  \tilde{x}.
    \end{equation}
Notice that in \eqref{eq:StabilityProof76}, $\boldsymbol{A}\left(\eta\right)$ and $\boldsymbol{\tilde{B}}(\eta)$ are affine in $\eta$.
Hence, in this case, proving the quadratic stability of \eqref{eq:StabilityProof76} implies robust stability; see \cite{becker1993quadratic} for proof.
The quadratic stability of \eqref{eq:StabilityProof76} is determined by the following inequalities:
\begin{equation}
\label{eq:thmProofEq1}
    \begin{aligned}
         \boldsymbol{A}(\eta)^\top \boldsymbol{Q} \boldsymbol{A}(\eta) - \boldsymbol{Q} &\prec 0, \;\; \forall \eta \in \Xi,\\
         \boldsymbol{Q} &\succ 0.
    \end{aligned}
\end{equation}
Since $\Xi$ is a convex set, inequality \eqref{eq:thmProofEq1} is satisfied if
\begin{equation}
\label{eq:thmProofEq2}
    \begin{aligned}
         \boldsymbol{A}(\xi^i)^\top \boldsymbol{Q} \boldsymbol{A}(\xi^i) - \boldsymbol{Q} &\prec 0, \\
         \boldsymbol{Q} &\succ 0,
    \end{aligned}
\end{equation}
holds for all vertices $\xi^i$ of $\Xi$.
% where $\xi^i \in \Xi$ for all $i \in \{1,...,p\}$.
Then, applying Schur complement,  \eqref{eq:thmProofEq2} yields
\begin{equation}
    \left[\begin{array}{cc}
        \boldsymbol{Q} &  \boldsymbol{Q}\boldsymbol{A}(\xi^i)\\
        \boldsymbol{A}(\xi^i)^\top \boldsymbol{Q} & \boldsymbol{Q}
    \end{array}\right] \succ 0, 
\end{equation}
for all vertices $\xi^i$ of $\Xi$.
\end{IEEEproof}

% ============================================================================================================================================
% ========================================================= Simulations Results ==============================================================
% ============================================================================================================================================

\section{Simulations Results}
\label{sec:mmcSimulationResults}

To validate the performance of the proposed controller, we study its closed-loop response, considering two scenarios: (i) an average equivalent circuit of the MMC simulated in PLECS/Simulink equivalent to \eqref{eq:mmcBilinearModelGlobal} and (ii) a linear average model equivalent to \eqref{eq:mmcLinearModelGlobal}.
In both cases, we use a direct AC/AC MMC to interface a medium-frequency transformer (MFT) for ultra-fast chargers with a medium-voltage grid, as in \cite{Pereira2021ExampleACAC}.
The MMC parameters are $C_{\gamma} {=}4$mF, $L_m{=}3$mH and $R_m{=}50$m$\Omega$, and Table~\ref{TabChap3:MMCSystemDescription} shows its nominal operation during vehicle-to-grid (V2G) mode. 
\begin{table}[htbp]
    \centering
    \begin{tabular}{|l|l|l|l|}
        \hline Nominal Value & Notation & Grid & MFT \\
        \hline Active Power &  $P^{g}_{m}$ &$1 \mathrm{MW}$ & $1 \mathrm{MW}$  \\
        % \hline Reactive Power & $Q^{g}_{m}$ & 0 VA & 0 VA \\
        \hline Voltage & $[\hat{\mathrm{V}}^g_m, \hat{\mathrm{V}}^z]$ & $25 \mathrm{kV}$ & $10 \mathrm{kV}$  \\
        \hline Current & $[\hat{\mathrm{I}}^{g*}_m, \hat{\mathrm{I}}^{z*}_m]$ & $80 \mathrm{A}$ & $101.15 \mathrm{A}$ \\
        \hline Frequency & $[f_1,f_2]$ & $50 \mathrm{Hz} $ & $1 \mathrm{kHz}$ \\
        \hline
    \end{tabular}
    \caption{Power requirements of the MMC in the Simulink environment.}
    \label{TabChap3:MMCSystemDescription}
\end{table}

In this case, we apply Theorem~\ref{lmm:ControllerSynthesis} to find a controller that maintains the MMC within a safe region of operation, i.e., $-0.1 \bar x_{\text{nom}} \leq  e_x \leq 0.1  \bar x_{\text{nom}}$ and  $-0.08 u_{\text{nom}}\leq e_u\leq 0.08 u_{\text{nom}}$, where $\bar x_{\text{nom}} = \hat{\mathrm{I}}^{g*}_m+\hat{\mathrm{I}}^{z*}_m$ and $u_{\text{nom}}=\hat{\mathrm{V}}^g_m+ \hat{\mathrm{V}}^z.$ The resulting controller with a structure as \eqref{eq:definitionMIMOController} yields
\begin{subequations}
\label{eq:resultingMIMOController}
    \begin{align}
        K_x &= \operatorname{diag}(K^a_x, K^b_x, K^c_x),\\
        K_w &= [\operatorname{diag}(K^{a}_{w}, K^{b}_{w}, K^{c}_{w}), [K^{z}_{w};K^{z}_{w};K^{z}_{w}]],
    \end{align}
\end{subequations}
where, for each phase $m {\in} \{a,b,c\}$,
\begin{equation*}
    K^m_x =-148.62\mathbf{I}_{2}, K^m_w =\left[\begin{array}{rr}
             -0.7621 & -0.0015   \\
              0.7621 & 0.0015
        \end{array}\right],  
\end{equation*}
and 
\begin{equation*}
    K^z_w =\left[\begin{array}{cc}
             2.4919 & -0.1902  \\
             2.4919 & -0.1902
        \end{array}\right].
\end{equation*}
% Moreover, by assessing the MMC parameters, we can confirm that Theorem~\ref{lmm:mmcStability} is satisfied.
Moreover, to assess the MMC stability, the solution from \eqref{eq:inequalityStabilityLMim} of Theorem~\ref{lmm:mmcStability} yields $\boldsymbol{Q} \succ 0$, i.e., $\boldsymbol{Q} =\operatorname{diag}\left(Q,Q,Q\right)$ with
\begin{equation}
   Q =\left[ \begin{array}{c c c c}
     1.582 & -0.437 & -0.449 & -0.437 \\
    -0.437 & 1.583 & -0.437 & -0.449 \\
    -0.449 & -0.437 & 1.583 & -0.437 \\
    -0.437 & -0.449 & -0.437 & 1.583 
    \end{array} \right].
\end{equation}

\begin{figure}[htbp]
\centerline{\includegraphics[width=\columnwidth]{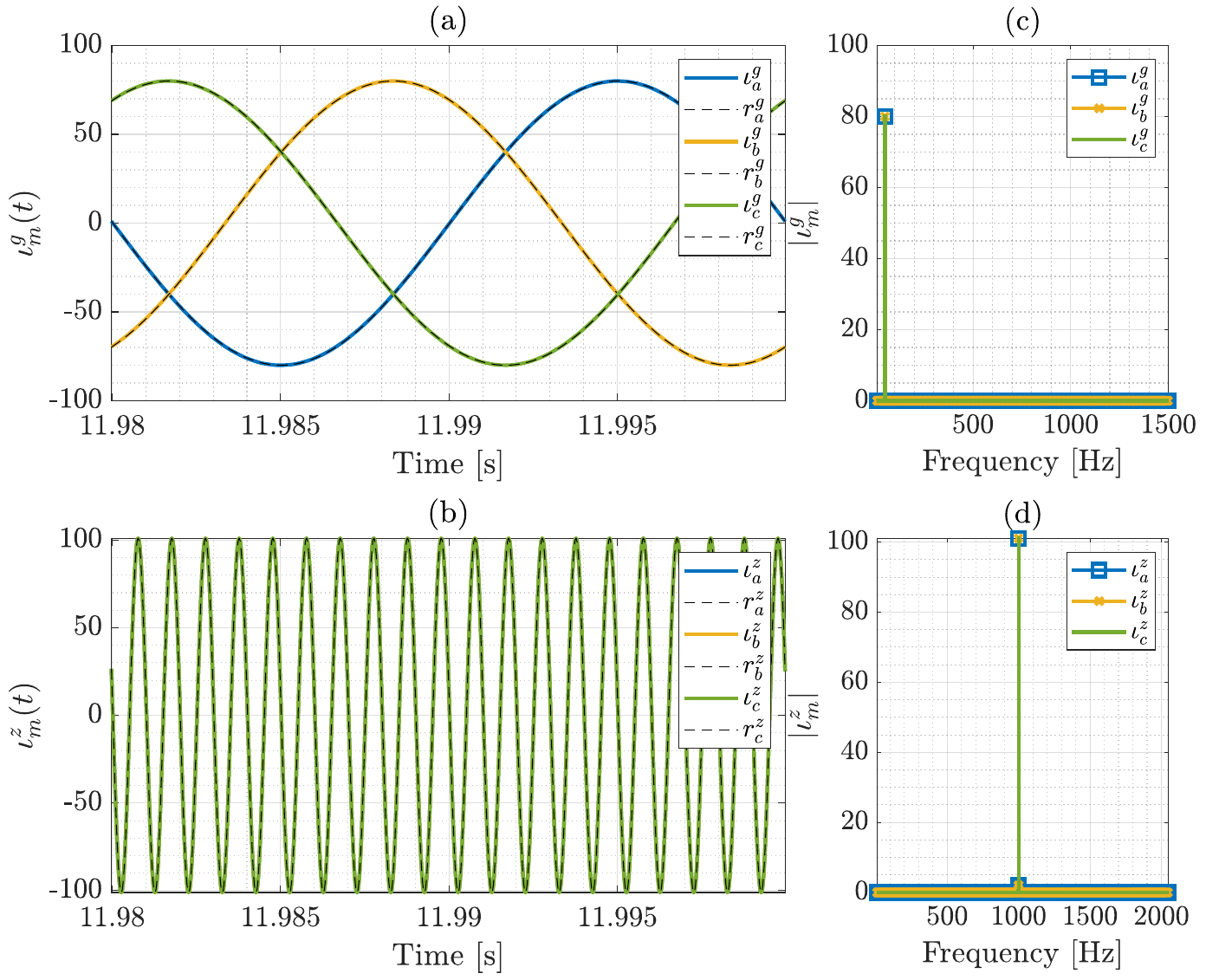}}
% This figure comes from the file: PlanPhD.drawio
% \centerline{\includesvg[width=\columnwidth]{Figures/2. Current Control Fig/CurrentControlPhDLandscape.svg}}
\caption{Simulation results from scenario (i): (a) Trajectory of the grid current, (b) trajectory of the output current, (c) FFT analysis of grid current, and (d) FFT analysis output current.}
\label{FigChap3:NonLinMMCOutputTracking}
\end{figure}

Figure~\ref{FigChap3:NonLinMMCOutputTracking} shows the simulation results corresponding to scenario (i).
In Figure~\ref{FigChap3:NonLinMMCOutputTracking}(a)-(b), we observe the tracked currents with the amplitude expected in Table~\ref{TabChap3:MMCSystemDescription}.
The FFT analysis in Figure~\ref{FigChap3:NonLinMMCOutputTracking}(c)-(d) shows a well-designed decoupling of the output currents, confirming the suitability of \eqref{eq:resultingMIMOController}.
Both currents, i.e., the grid current and the output current, have only visible components at their corresponding fundamental frequencies $f_1$ and $f_2$, respectively.
% Hence, it 

% 

\begin{figure}[htbp]
\centerline{\includegraphics[width=\columnwidth]{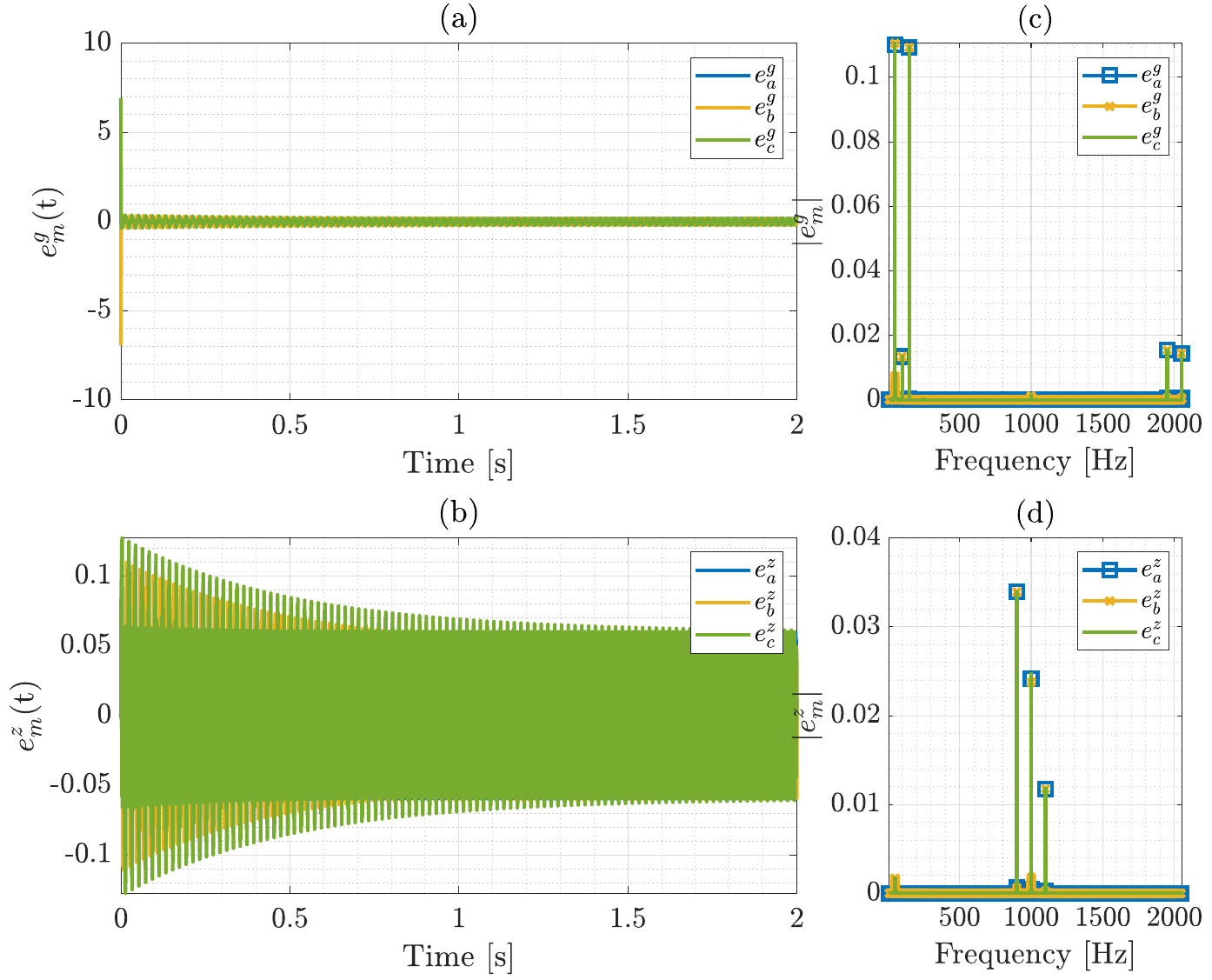}}
% This figure comes from the file: PlanPhD.drawio
% \centerline{\includesvg[width=\columnwidth]{Figures/2. Current Control Fig/CurrentControlPhDLandscape.svg}}
\caption{Simulation results from scenario (i): (a) Error of the grid current $e^g_m$, (b) Error of the output current $e^z_m$, (c) FFT analysis of grid current error, and (d) FFT analysis output current error.}
\label{FigChap3:NonLinMMCErrorTracking}
\end{figure}

Figure~\ref{FigChap3:NonLinMMCErrorTracking} shows the deviation of the MMC currents as \eqref{eq:currentControlProblem}, i.e., $\varepsilon_{\{1,2,5,6,9,10\}}$, resulting from scenario (i).
As expected from Theorem~\ref{lmm:mmcStability}, the deviation is stable and bounded; see Figure~\ref{FigChap3:NonLinMMCErrorTracking}(a)-(b).
The FFT analysis of the tracking error shows some small components at 50Hz and 1kHZ corresponding to the grid and output current, respectively.
\textcolor{black}{Hence, Figure~\ref{FigChap3:NonLinMMCErrorTracking}(c)-(d) also confirms the good alignment to the grid and the MFT phases.}
We also observe some components in other frequencies, but the amplitude is relatively small, i.e., $< 0.125\%$.

\begin{figure}[htbp]
\centerline{\includegraphics[width=\columnwidth]{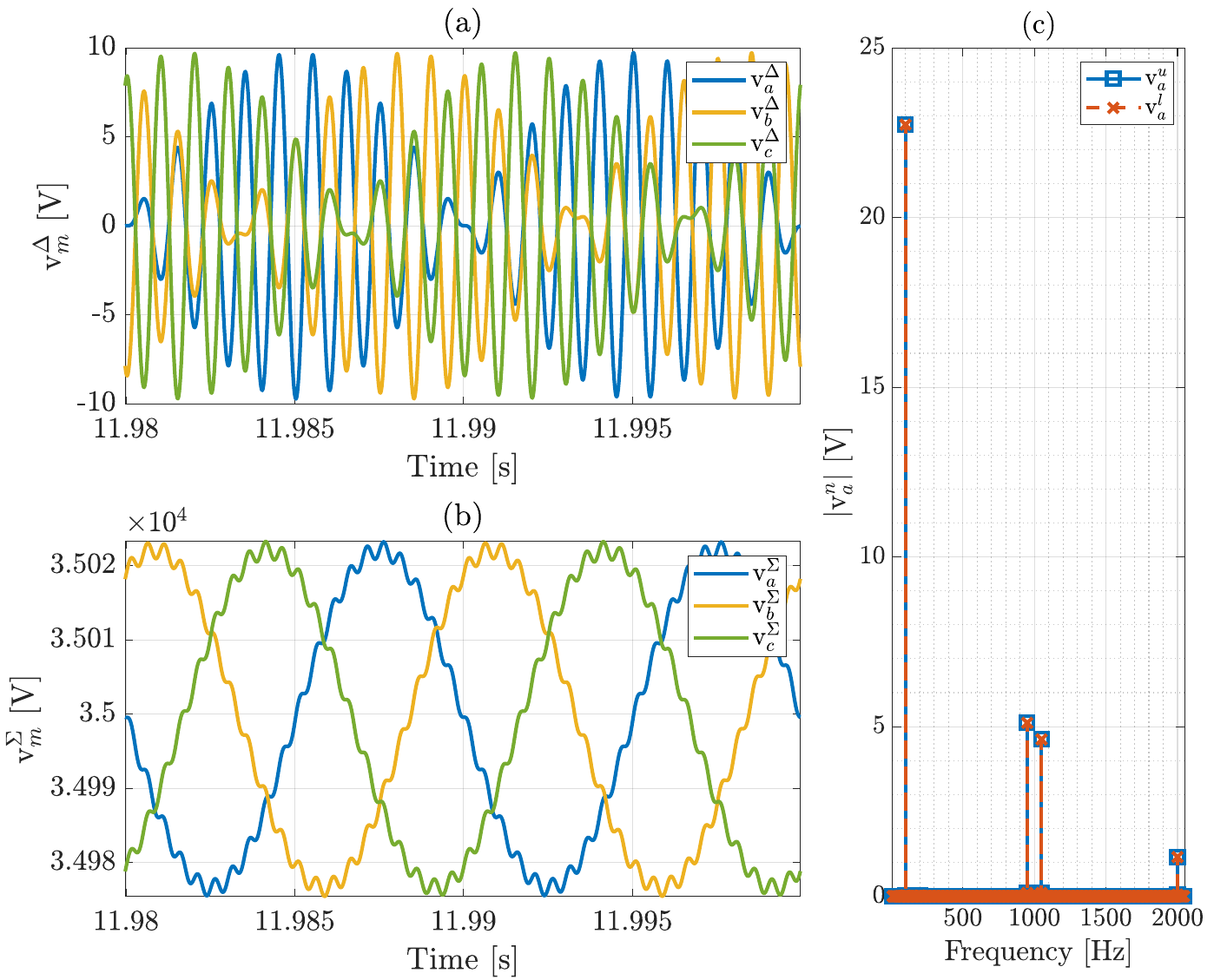}}
% This figure comes from the file: PlanPhD.drawio
% \centerline{\includesvg[width=\columnwidth]{Figures/2. Current Control Fig/CurrentControlPhDLandscape.svg}}
\caption{Simulation results from scenario (i): (a) Trajectory of the differential-mode total arm voltage, (b) trajectory of the common-mode total arm voltage, (c) FFT analysis of the total arm voltage of phase a.}
\label{FigChap3:NonLinMMCStoredVoltageTracking}
\end{figure}

Figure~\ref{FigChap3:NonLinMMCStoredVoltageTracking} depicts the behaviour of the total arm voltage when scenario (i) was simulated.
In Figure~\ref{FigChap3:NonLinMMCStoredVoltageTracking}(a)-(b), the total arm voltage in the differential and common modes shows that the MMC is operating stably.
The FFT analysis in Figure~\ref{FigChap3:NonLinMMCStoredVoltageTracking}(c) proves that the total arm voltage ripples ($\leq 0.1\%$) have components in the expected frequencies, as it was explained in \cite{Pereira2023ExampleACAC}.

\begin{figure}[htbp]
\centerline{\includegraphics[width=\columnwidth]{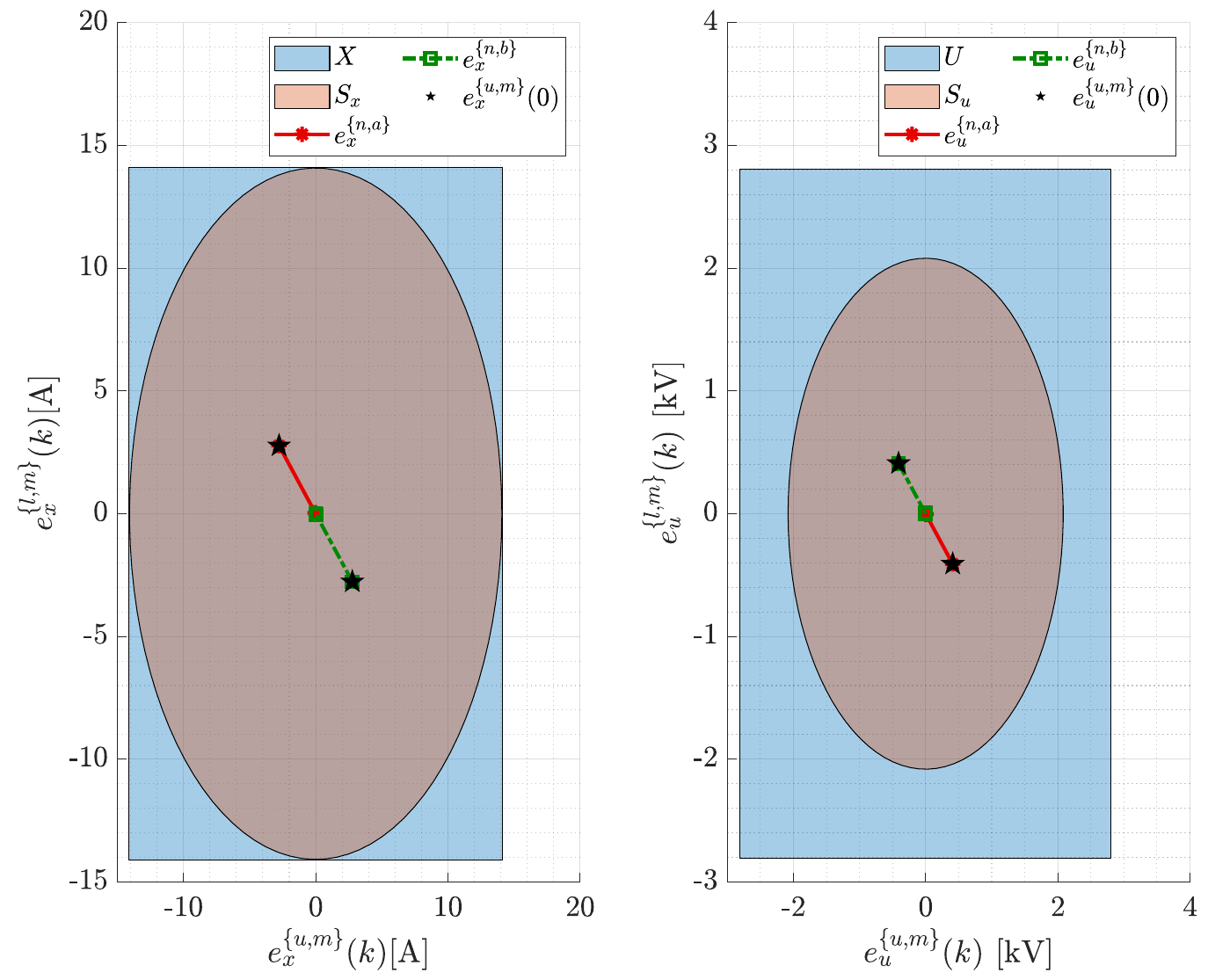}}
% This figure comes from the file: PlanPhD.drawio
% \centerline{\includesvg[width=\columnwidth]{Figures/2. Current Control Fig/CurrentControlPhDLandscape.svg}}
\caption{Simulation results from scenario (ii) with initial condition $(\star \; e(0))$: (a) Trajectory of the state tracking error $e_x$, operation region $\mathbb{S}_x$ and constraint $\mathbb{X}$; (b) trajectory of the control input tracking error  $e_u$, operation region $\mathbb{S}_u$ and constraint $\mathbb{U}$.}
\label{FigChap3:LinMMCStateErrorTracking}
\end{figure}

We evaluate the design of the controller through the state tracking error dynamics corresponding to the simulated scenario (ii). 
For simplicity, in Figure~\ref{FigChap3:LinMMCStateErrorTracking}, we plot the projection of $\mathbb{S}_x$ corresponding to phases b and c. Note that for all $k \in \mathbb{N}$, $e_x(k) \in \mathbb{S}_x \subseteq \mathbb{X}$ and $e_u(k) \in \mathbb{S}_u \subseteq \mathbb{U}$.

% ============================================================================================================================================
% =======================================================  Experimental Verification  ========================================================
% ============================================================================================================================================

\section{Experimental Verification}
\label{sec:ExperimentalVerification}
\begin{figure}[htbp]
\centerline{\includegraphics[width=0.8\columnwidth]{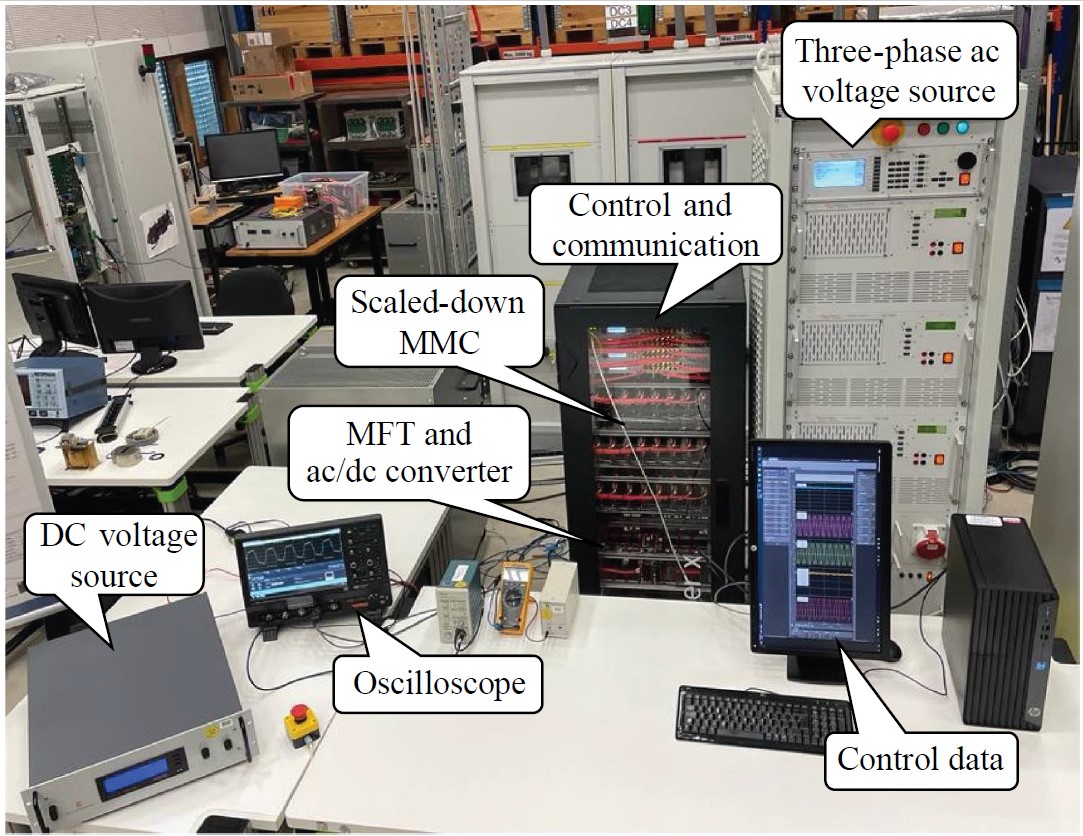}}
% This figure comes from the file: PlanPhD.drawio
% \centerline{\includesvg[width=\columnwidth]{Figures/2. Current Control Fig/CurrentControlPhDLandscape.svg}}
\caption{Photograph of the experimental setup with the scaled-down MMC.}
\label{FigChap3:Photograph}
\end{figure}

The proposed controller is verified with a scaled-down prototype, shown in Figure~\ref{FigChap3:Photograph}. 
The experimental setup implements an AC/AC MMC to interface a three-phase voltage source, i.e., the grid simulator, with a medium-frequency transformer (MFT) primary side.
The MMC is composed of PEH2015 and PEB8038 modules from Imperix. 
The system is controlled using four Imperix B-Box RCP 3.0 fast prototyping control platform modules, which run the control scheme described in Section~\ref{sec:mmcCurrentController} in discrete-time implementation. 
In addition, the data acquired via the controllers has a sample rate ($T_s$) of 50 kHz.
The MMC parameters are $N{=}{4}$, $C_{\gamma} {=}5$mF, $L_m{=}2.36$mH and $R_m{=}50$m$\Omega$ and Table~\ref{TabChap3:MMCSystemDescription2} shows its nominal operation during vehicle-to-grid (V2G) mode. 

\begin{table}[htbp]
    \centering
    \begin{tabular}{|l|l|l|l|}
        \hline Nominal Value & Notation & Grid & MFT \\
        \hline Active Power &  $P^{g}_{m}$ &$1/3~\mathrm{kW}$ & $1/3~\mathrm{kW}$  \\
        % \hline Reactive Power & $Q^{g}_{m}$ & 0 VA & 0 VA \\
        \hline Voltage & $[\hat{\mathrm{V}}^g_m, \hat{\mathrm{V}}^z]$ & $300 \mathrm{V}$ & $150 \mathrm{V}$  \\
        \hline Current & $[\hat{\mathrm{I}}^{g*}_m, \hat{\mathrm{I}}^{z*}_m]$ & $3.33 \mathrm{A}$ & $3.395 \mathrm{A}$ \\
        \hline Frequency & $[f_1,f_2]$ & $50 \mathrm{Hz} $ & $1 \mathrm{kHz}$ \\
        \hline
    \end{tabular}
    \caption{Power requirements of the AC/AC scaled-down prototype.}
    \label{TabChap3:MMCSystemDescription2}
\end{table}

In this case, we apply Theorem~\ref{lmm:ControllerSynthesis} to find a controller that maintains the MMC within a safe region of operation, i.e., $-0.1 \bar x_{\text{nom}} \leq  e_x \leq 0.1  \bar x_{\text{nom}}$ and  $-0.08 u_{\text{nom}}\leq e_u\leq 0.08 u_{\text{nom}}$, where $\bar x_{\text{nom}} = \hat{\mathrm{I}}^{g*}_m+\hat{\mathrm{I}}^{z*}_m$ and $u_{\text{nom}}=\hat{\mathrm{V}}^g_m+ \hat{\mathrm{V}}^z.$ The resulting controller with a structure as \eqref{eq:definitionMIMOController} yields
\begin{subequations}
\label{eq:resultingMIMOController2}
    \begin{align}
        K_x &= \operatorname{diag}(K^a_x, K^b_x, K^c_x),\\
        K_w &= [\operatorname{diag}(K^{a}_{w}, K^{b}_{w}, K^{c}_{w}), [K^{z}_{w};K^{z}_{w};K^{z}_{w}]],
    \end{align}
\end{subequations}
where, for each phase $m {\in} \{a,b,c\}$,
\begin{equation*}
    K^m_x =-8.9465\mathbf{I}_{2}, K^m_w =\left[\begin{array}{rr}
             -0.95 & -0.0040   \\
              0.95 & 0.0040
        \end{array}\right],  
\end{equation*}
and 
\begin{equation*}
    K^z_w =\left[\begin{array}{cc}
             1.1835 & -0.3265  \\
             1.1835 & -0.3265
        \end{array}\right].
\end{equation*}
% Moreover, by assessing the MMC parameters, we can confirm that Theorem~\ref{lmm:mmcStability} is satisfied.
Moreover, to assess the MMC stability, the solution from \eqref{eq:inequalityStabilityLMim} of Theorem~\ref{lmm:mmcStability} yields $\boldsymbol{Q} \succ 0$, i.e., $\boldsymbol{Q} =\operatorname{diag}\left(Q,Q,Q\right)$ with
\begin{equation}
   P =\left[ \begin{array}{c c c c}
     1.582 & -0.437 & -0.448 & -0.437 \\
    -0.437 & 1.582 & -0.437 & -0.449 \\
    -0.448 & -0.437 & 1.582 & -0.437 \\
    -0.437 & -0.448 & -0.437 & 1.582 
    \end{array} \right].
\end{equation}

\begin{figure}[htbp]
\centerline{\includegraphics[width=\columnwidth]{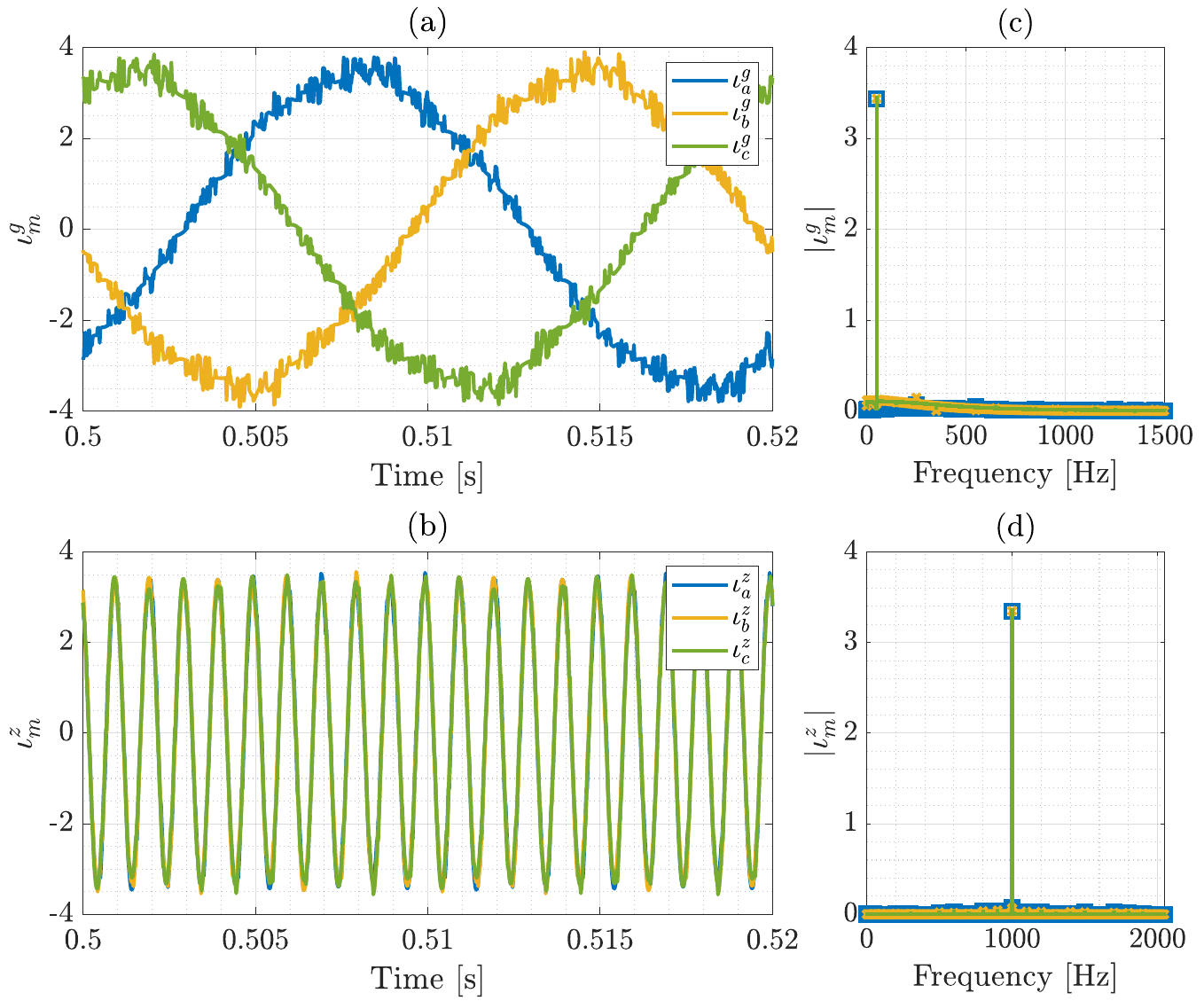}}
% This figure comes from the file: PlanPhD.drawio
% \centerline{\includesvg[width=\columnwidth]{Figures/2. Current Control Fig/CurrentControlPhDLandscape.svg}}
\caption{(a) Steady-state trajectory of the grid currents, (b) Steady-state trajectory of the transformer currents, (c) FFT analysis of grid currents, and (d) FFT analysis transformer currents.}
\label{FigChap3:ImperixMMCOutputTracking}
\end{figure}

Figure~\ref{FigChap3:ImperixMMCOutputTracking} shows steady-state dynamics of the MMC output currents.
By assessing Figure~\ref{FigChap3:ImperixMMCOutputTracking}(a)-(b) and Figure~\ref{FigChap3:ImperixMMCOutputTracking}(c)-(d), we observe that the tracked currents have the amplitude expected in Table~\ref{TabChap3:MMCSystemDescription}.
As in previous examples, the FFT analysis in Figure~\ref{FigChap3:ImperixMMCOutputTracking}(c)-(d) shows a well-designed decomposition of the output currents, validating the suitability of \eqref{eq:resultingMIMOController}.
% Both currents, i.e., the grid current and the output current, have only visible components at their corresponding fundamental frequencies $f_1$ and $f_2$, respectively.

\begin{figure}[htbp]
\centerline{\includegraphics[width=\columnwidth]{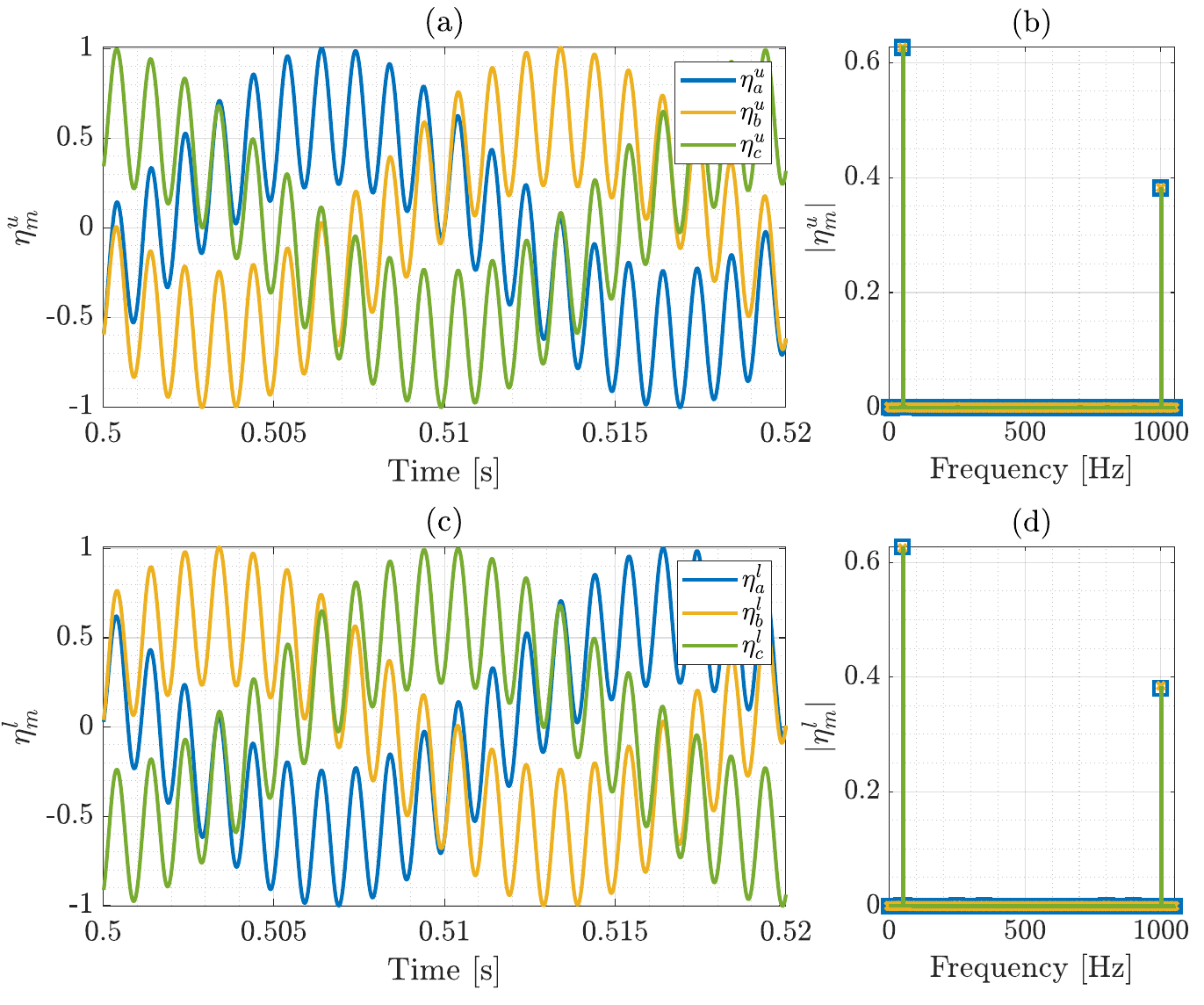}}
% This figure comes from the file: PlanPhD.drawio
% \centerline{\includesvg[width=\columnwidth]{Figures/2. Current Control Fig/CurrentControlPhDLandscape.svg}}
\caption{(a) Steady-state trajectory of the insertion index of the upper arms, (b) Steady-state trajectory of the insertion index of the lower arms, (c) FFT analysis of the insertion index of the upper arms, and (d) FFT analysis of the insertion index of the lower arms.}
\label{FigChap3:ImperixMMCControlInputTracking}
\end{figure}

These output currents are generated by the insertion index presented in Figure~\ref{FigChap3:ImperixMMCControlInputTracking}.
\textcolor{black}{In this case, the insertion index remains between the intended operating region, while providing a good performance.}
Additionally, the FFT shows only components at the desired frequencies, i.e., $f_1$ and $f_2$.
The amplitude corresponding to these frequencies also follows the MMC ideal dynamics expressed in \cite{Pereira2021ExampleACAC}.

% In this section, we assess the suitability of the controller \eqref{eq:resultingMIMOController2} by analyzing the control input dynamics, i.e., arm voltage ($u^n_m$), in the differential and common modes. This differential and common modes are defined as $u^{\Delta}_m = 0.5(u^u_m-u^l_m)$ and $u^{\Sigma}_m = 0.5(u^u_m+u^l_m)$. Note that the ideal operation of the MMC corresponds to $|u^{\Delta}_m| \approx |v^g_m|$ and  $|u^{\Sigma}_m| \approx |v^z|$, as explained in  \cite{Pereira2021ExampleACDC}. 
% Figure~\ref{FigChap3:ImperixMMCControlInputTracking} shows the arm voltage dynamics.
% The FFT analysis in Figure~\ref{FigChap3:ImperixMMCControlInputTracking}(c)-(d) confirms the well-designed decoupling of both sides of the MMC. 

% \begin{figure}[htbp]
% \centerline{\includegraphics[width=\columnwidth]{Figure/ImperixMMCControlInputTracking.pdf}}
% % This figure comes from the file: PlanPhD.drawio
% % \centerline{\includesvg[width=\columnwidth]{Figures/2. Current Control Fig/CurrentControlPhDLandscape.svg}}
% \caption{(a) Steady-state trajectory of the differential-mode arm voltage, (b) Steady-state trajectory of the common-mode arm voltage, (c) FFT analysis of the differential-mode arm voltage, and (d) FFT analysis of the common-mode arm voltage.}
% \label{FigChap3:ImperixMMCControlInputTracking}
% \end{figure}

\begin{figure}[htbp]
\centerline{\includegraphics[width=\columnwidth]{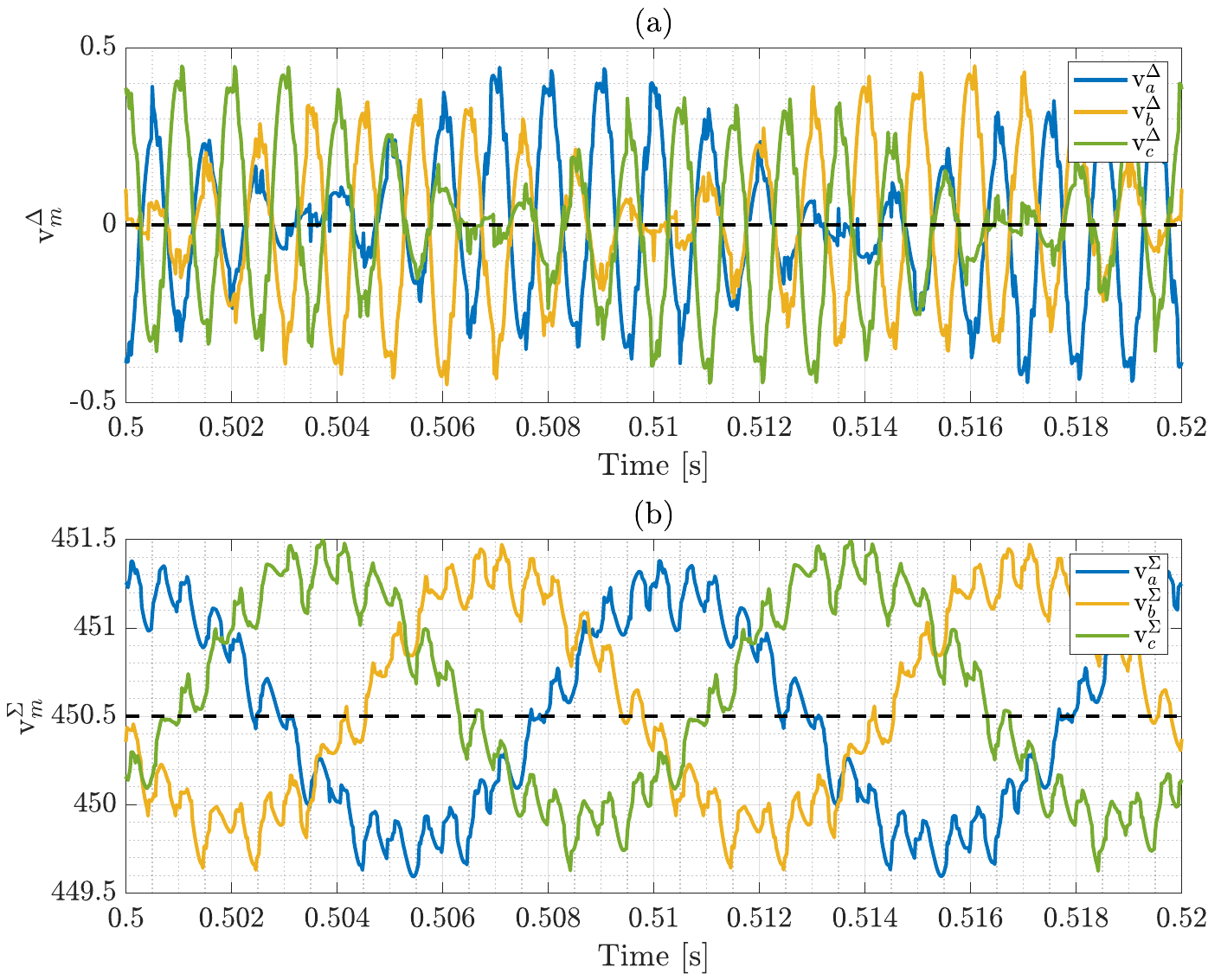}}
% This figure comes from the file: PlanPhD.drawio
% \centerline{\includesvg[width=\columnwidth]{Figures/2. Current Control Fig/CurrentControlPhDLandscape.svg}}
\caption{(a) Steady-state trajectory of the differential-mode total arm voltage, (b) Steady-state trajectory of the common-mode total arm voltage,}
\label{FigChap3:ImperixMMCVoltageTracking}
\end{figure}

Figure~\ref{FigChap3:ImperixMMCVoltageTracking} depicts the behaviour of the differential and common modes of the total arm voltage.
Note that the black dashed lines represent the average value of both signals.
In this case, as in \eqref{eq:sotredVoltagetControlProblem}, the average value of the differential mode is equal to zero. 
The average of the common mode is not exactly $ \hat{\mathrm{V}}^g_m + \hat{\mathrm{V}}^z$ as in \eqref{eq:sotredVoltagetControlProblem}.
This was expected due to the model mismatching, yet it does not compromise the stable and efficient operation of the MMC.

% ============================================================================================================================================
% ===========================================================  Conclusions  ==================================================================
% ============================================================================================================================================

\section{Conclusion}
\label{sec:Conclusion}
In this paper, we propose a controller for tracking the MMC currents while keeping the stored capacitor voltage stable. 
As an outcome of this paper, we provide evidence that a synthesized feedback gain is suitable for controlling the MMC.
The static feedback gain is designed to achieve the required closed-loop performance while maintaining stability and safety.
Moreover, the complexity of the proposed control scheme is simpler since no PLL is needed.
Evidence of the advantage of the proposed controller is shown in different scenarios (i) linear average model, (ii) average equivalent MMC circuit, and (iii) prototype scaled-down MMC.

% ============================================================================================================================================
% =============================================================  Appendix  ===================================================================
% ============================================================================================================================================

\appendices
\section*{Acknowledgment}
The authors would like to thank Jorge L. Duarte and Korneel G. E. Wijnands for sharing the topology of the modular multilevel converter for ultra-fast charging stations and for their useful comments.

\section*{Appendixes}
\subsection{State and Control Input Error Dynamics}
\label{appx:DerivationErrorDynamics}
Recalling \cite[Theorem 1 c.f.]{Francis1976LOT}, let assume that there exist $\Pi$ and $\Gamma$ such that 
\begin{equation}
\label{eq:Derivation1}
\begin{aligned}
    & \Pi S=A \Pi+B \Gamma+E, \\
    & C \Pi=O.
\end{aligned}  
\end{equation}
Then, by multiplying the exogenous input vector, \eqref{eq:Derivation1} yields
\begin{equation}
\label{eq:Derivation2}
    \begin{aligned}
    & \Pi S w =A \Pi w+B \Gamma w +E w, \\
    & C \Pi w=O w.
\end{aligned} 
\end{equation}
Note that \eqref{eq:Derivation2} confirms that defining the steady-state as
\begin{equation}
\label{eq:Derivation3}
    x^{ss} := \Pi w \quad   \text{and} \quad u^{ss} := \Gamma w
\end{equation} 
guarantees that the tracking error is equal to zero, i.e.,
\begin{equation}
\label{eq:Derivation3}
        e(k) =  C x(k) - O w(k) = 0.
\end{equation}
Equation \eqref{eq:Derivation2} can be re-written as
\begin{equation}
    \label{eq:Derivation4}
    \Pi {w} (k+1)  =A \Pi w+B \Gamma w+E w        
\end{equation}
and, consequently, as
\begin{equation}
\label{eq:Derivation5}
    {x}^{{ss}} (k+1) =A x^{ss}+B x^{ss}+E w.
\end{equation}

By defining $e_x=x-x^{ss},$ and $ e_u=u-u^{ss}$, it follows immediately that
\begin{equation}
\label{eq:Derivation6}
\begin{aligned}
e_x (k+1) &=  x(k+1) - x^{ss}(k+1), \\
          &= A x(k) + B u(k) + E w(k)  -  \Pi S w. 
        \end{aligned}
    \end{equation}
Substituting \eqref{eq:definitionMIMOController} in \eqref{eq:Derivation6}, it yields    
\begin{equation}
\label{eq:Derivation7}
        e_x (k{+}1){=} (A {+} B K_x ) x {+} (B K_w ) w {-} (A \Pi {+} B \Gamma ) w.
\end{equation}
Rearranging \eqref{eq:Derivation7} the error state dynamics is defined by an autonomous system, i.e.,
\begin{equation}
\label{eq:Derivation8}
\begin{aligned}
    e_x (k{+}1) &{=} (A {+} B K_x ) (e_x{ +} \Pi w ) {+} \\  & \;\;\; (B K_w ) w {-} (A \Pi {+} B (K_w {+} K_x\Pi))  w\\
                &{=}  A e_x(k) {+ } B K_x e_x(k).
\end{aligned}
\end{equation}    
From \eqref{eq:Derivation8}, the control input corresponding to state error yields $e_u := K_x e_x$, i.e., 
\begin{equation}
\label{eq:Derivation9}
    e_u = K_x (x - x^{ss}).
\end{equation}
Note that \eqref{eq:Derivation9} is equivalent to the control input error definition, i.e., $e_u = u - u^{ss}$ as in \eqref{eq:scontrolInptErrorDefnition}, since $u^{ss} = \Gamma w$ and
\begin{equation}
    u =  K_x x + \underbrace{\left(\Gamma-K_x \Pi \right)}_{K_w} w.
\end{equation}

% ============================================================================================================================================
% =============================================================  References  ===================================================================
% ============================================================================================================================================

% \vspace{-0.4cm}
% \bibliography{main_bib}
\bibliography{bibliography/mybibfile}
% \vspace{-0.4cm}
\bibliographystyle{IEEEtran}

\end{document}